\documentclass[11pt]{article}

\usepackage{amsfonts,amsmath}

\begin{document}
\date{}

\title{\textbf{Regularization of odd-dimensional AdS gravity: Kounterterms. }%
}
\author{Rodrigo Olea \medskip \medskip \\
{\small \textit{Centro Multidisciplinar de Astrof\'{\i}sica -
CENTRA,
Departamento de F\'{\i}sica,}}\\
{\small \textit{Instituto Superior T\'{e}cnico, Universidade
T\'{e}cnica de
Lisboa,}} \\
{\small \textit{Av. Rovisco Pais 1, 1049-001 Lisboa, Portugal.}}\\
{\small E-mail: rolea@fisica.ist.utl.pt}}
\maketitle

\begin{abstract}
As an alternative to the standard Dirichlet counterterms
prescription, I introduce the concept of \emph{Kounterterms} as the
boundary terms with explicit dependence on the extrinsic curvature
$K_{ij}$ that regularize the AdS gravity action. A suitable choice
of the boundary conditions --compatible with any asymptotically AdS
(AAdS) spacetime-- ensures a finite action principle for all odd
dimensions. Background-independent conserved quantities are obtained
as Noether charges associated to asymptotic symmetries and their
general expression appears naturally split in two parts. \ The first
one gives the correct mass and angular momentum for AAdS black holes
and vanishes identically for globally AdS spacetimes. Thus, the
second part is a covariant formula for the vacuum energy in AAdS
spacetimes and reproduces the results obtained by the Dirichlet
counterterms method in a number of cases. It is also shown that this
Kounterterms series regularizes the Euclidean action and recovers
the correct black hole thermodynamics in odd dimensions.
\end{abstract}

\section{Introduction}

 In the context of AdS/CFT correspondence \cite{Maldacena},
Witten sketched the program to regularize the action for AdS
spacetimes \cite{Witten}, which was carried out in detail by
Hennigson and Skenderis in Ref.\cite{Sk-He}.

This procedure, known as holographic renormalization, considers a
generic form of the metric for an asymptotically AdS (AAdS)
spacetime

\begin{equation}
ds^{2}=\mathcal{G}_{\mu \nu }dx^{\mu}dx^{\nu}=\frac{\ell ^{2}}{4\rho ^{2}}%
d\rho ^{2}+\frac{g_{ij}\left( \rho ,x\right) }{\rho }dx^{i}dx^{j}
\label{confcoord}
\end{equation}%
where $\rho $ is the radial coordinate of a manifold $M$, whose
boundary is located at $\rho =0$, and $\ell $ is the AdS radius.
This coordinates choice is suitable to describe the conformal
structure of the boundary, whose metric $g_{ij}\left( \rho
,x\right) $ accepts a regular expansion \cite{F-G}

\begin{equation}
g_{ij}\left( \rho ,x\right) =g_{(0)ij}\left( x\right) +\rho g_{(1)ij}\left(
x\right) +\rho ^{2}g_{(2)ij}\left( x\right) +...  \label{FGexp}
\end{equation}%
where $g_{(0)ij}$ is a given initial value for the metric.

Solving the Einstein equations in this frame reconstructs the
spacetime from the boundary data, determining the coefficients
$g_{(k)}$ as covariant functionals in the boundary metric $g_{(0)}$
(that contain $k$ derivatives of $x^{i}$). Then, in order to
preserve general covariance at the boundary, it is necessary to
invert the series to express all the quantities as functions of the
boundary metric $h_{ij}=g_{ij}/\rho $ \cite{dHSS}.

In this way, the counterterms method proposes a regularization
scheme that consists in the addition to the Dirichlet action of
local functionals of the boundary metric $h_{ij}$, the intrinsic
curvature $R_{ij}^{kl}$ of the boundary and covariant derivatives
of the boundary Riemann $\nabla _{m}R_{ij}^{kl}$. In $D=d+1$
dimensions the regularized action reads
\begin{equation}
I=-\frac{1}{16\pi G}\int_{M}d^{d+1}x\sqrt{-\mathcal{G}}\left( \hat{R}%
-2\Lambda \right) -\frac{1}{8\pi G}\int_{\partial M}d^{d}x\sqrt{-h}%
K+\int_{\partial M}d^{d}x\mathcal{B}_{ct}(h,R,\nabla R).  \label{Ict}
\end{equation}%

In the above action, hatted curvatures refer to
$(d+1)-$dimensional ones, the cosmological constant is $\Lambda
=-d(d-1)/2\ell ^{2}$ and $K=K_{ij}h^{ij}$ is the trace of the
extrinsic curvature.

The Gibbons-Hawking term ensures a well-posed variational
principle for a Dirichlet boundary condition on the metric
$h_{ij}$ \cite{Gibbons-Hawking}, that is still valid in presence of the
Dirichlet counterterms series because its functional variation is expressed as
$\delta \mathcal{B}_{ct}=(\delta \mathcal{B}_{ct}/\delta h_{ij})\delta h_{ij}$.

As a consequence, a regularized stress tensor for AdS spacetimes is
obtained \cite{Ba-Kr,Emparan-Johnson-Myers} using the Brown-York
quasilocal energy-momentum tensor definition \cite{Brown-York},
without reference to any background solution.

A novel feature of this approach is the appearance --in $D=2n+1$
dimensions-- of a vacuum energy for AdS spacetime, that is clearly
unobservable in background-dependent methods. In five dimensions,
the matching between this vacuum energy and the Casimir energy
induced by a precise boundary CFT ($\mathcal{N}=4$ SYM theory with
gauge group $SU(N)$) is one of the best known examples of the
AdS/CFT correspondence \cite{Ba-Kr}.

Despite the fact this regularization procedure provides a
systematic way to construct the Dirichlet counterterms series, in
practice, the number of possible counterterms increase drastically
with the dimension. Even for a given dimension, the finiteness of
the conserved charges for a more complex solution would require a
significant addition of counterterms respect to the same problem,
for instance, in Schwarzschild-AdS black hole. Moreover, these
extra terms do not seem to obey any particular pattern
\cite{Das-Mann}.

In recent papers \cite{Pa-Sk1,Pa-Sk2}, the problem of Dirichlet
counterterms in AdS gravity has been reformulated as an
initial-value problem for the extrinsic curvature. This results in
a simpler algorithm to obtain the series $\mathcal{B}_{ct}$ but,
however, the full series for an arbitrary dimension is still
unknown.

Furthermore, in the counterterms method is not clear where the
vacuum energy is coming from, i.e., which boundary terms are
responsible for the shifting of the zero-point energy of AdS or
whether it can be obtained from a general covariant formula for
any AAdS spacetime.

Therefore, a natural question arises: Is there another
counterterms series that also regularizes the Einstein-Hilbert
action with negative cosmological constant, and whose expression
can be worked out in any dimension?

Certainly the answer to that problem implies the departure from a
standard action principle based on a Dirichlet boundary condition on
the metric $h_{ij}$.

Such construction may at first sound too ambitious. However, we have
evidence coming from even-dimensional AdS gravity, where considering
a boundary condition for the spacetime curvature
\begin{equation}
\hat{R}_{\alpha \beta }^{\mu \nu }+\frac{1}{\ell ^{2}}\delta _{\lbrack
\alpha \beta ]}^{[\mu \nu ]}=0  \label{Reqdelta}
\end{equation}%
on $\partial M$ has been a good alternative to produce a finite
action principle \cite{ACOTZ4,ACOTZ2n}. In that case, the gravity
action is supplemented by the Euler term $\mathcal{E}_{2n}$ with a
coupling constant fixed demanding this generic asymptotic condition.

In $D=2n$ dimensions, the Euler theorem states that the Euler term $\mathcal{%
E}_{2n}$ is equivalent to the boundary term $B_{2n-1}$ (the $n-$th Chern
form) up to the Euler characteristic $\chi _{2n}$, a topological number for
the manifold $M$. From a dynamic point of view, $\chi _{2n}$ is just an
integration constant and then, the formula for the conserved charges is the
same if we supplement the Einstein-Hilbert-AdS action either with the
boundary term $B_{2n-1}$ or the Euler (bulk) term $\mathcal{E}_{2n}$ \cite%
{OleaJHEP}. In that sense, the regularizing effect of the Dirichlet
counterterm series can be replaced by the Euler term in even
dimensions. The clear advantage of this procedure is that we do not
need to perform any particular expansion of the metric, nor solving
the asymptotic equations in that frame to find the coefficients
$g_{(k)ij}=g_{(k)ij}(g_{(0)})$, nor inverting the series to express
all the quantities as covariant functionals of the boundary metric
$h_{ij}$. This procedure contrasts with the simplicity of fixing a
single global factor in $\mathcal{E}_{2n}$, given by the asymptotic
condition (\ref{Reqdelta}). In other words, considering the Euler
term as a single entity, its coupling constant comes from fixing the
leading order in the curvature,
because in the expansion (\ref{confcoord}),(\ref{FGexp}), the asymptotic Riemann reads%
\begin{eqnarray}
\hat{R}_{ij}^{k\rho }(\mathcal{G}) &=&O(\rho ^{2}), \\
\hat{R}_{i\rho }^{k\rho }(\mathcal{G}) &=&-\frac{1}{\ell ^{2}}\delta
_{i}^{k}+O(\rho ^{2}), \\
\hat{R}_{ij}^{kl}(\mathcal{G}) &=&-\frac{1}{\ell ^{2}}\delta _{\lbrack
ij]}^{[kl]}+\rho \left( R_{ij}^{kl}(g_{(0)})+\frac{1}{\ell ^{2}}\left(
\delta _{\lbrack i}^{k}g_{(1)j]}^{l}+g_{(1)[i}^{k}\delta _{j]}^{l}\right)
\right) +O(\rho ^{2}),
\end{eqnarray}%
where $g_{(1)j}^{i}=g_{(0)}^{il}g_{(1)lj}$.

In sum, what is remarkable in this approach is the fact that we have a
closed expression for the boundary term $B_{2n-1}$ in all even dimensions
and also, a deep connection between the regularizing boundary terms and
topological invariants.

The same approach cannot be applied to odd-dimensional AdS gravity, because
there are no topological invariants of the Euler class in $D=2n+1$, what
makes gravity in even and odd dimensions quite different. This is not so
surprising, because also in standard holographic renormalization there
appear technical differences respect the even-dimensional case: the
expansion (\ref{FGexp}) requires a $\log \rho $ term at order $\rho ^{n}$ to
be consistent with the equations of motion, the existence of Weyl anomaly,
the appearance of a vacuum energy for AdS space, etc.

Following a different strategy, a mechanism to regularize the AdS
gravity action in odd dimensions was proposed in \cite{MOTZodd}. A
boundary condition for the extrinsic curvature $K_{i}^{j}$ in AAdS
spaces is the key assumption that leads to a well-posed action
principle for a given boundary term $B_{2n}$. This asymptotic
condition was motivated by a similar construction in
Chern-Simons-AdS gravity \cite{MOTZCS}.

Indeed, in the three-dimensional case, where Einstein-Hilbert-AdS gravity
action is a Chern-Simons form for the group $SO(2,2)$, this prescription
regularizes the Euclidean action and the Noether charges with a single
boundary term that is one half the Gibbons-Hawking term. It can be proved
that a suitable expansion reduces the problem to the Dirichlet formulation
plus a topological invariant of the boundary metric \cite{Mis-Ole}.

At this point, as an alternative to the standard counterterms approach, we
introduce the concept of \emph{Kounterterms} as the boundary terms that
regularize the AdS gravity action, that posses explicit dependence on the
extrinsic curvature $K_{ij}$ and whose construction is based on boundary
conditions compatible with the Fefferman-Graham form of the metric (\ref%
{FGexp}).

In this article, we show the general tensorial form the Kounterterms adopt
in the odd-dimensional case.

\section{Kounterterms}

We consider the Einstein-Hilbert action with negative cosmological constant,
with the addition of a boundary term $B_{d}$%
\begin{equation}
I=-\frac{1}{16\pi G}\int_{M}d^{d+1}x\sqrt{-\mathcal{G}}\left( \hat{R}%
-2\Lambda \right) +c_{d}\int_{\partial M}d^{d}xB_{d}  \label{IgD}
\end{equation}%
instead of the standard Gibbons-Hawking term plus the counterterms
series. Here, $c_{d}$ is a coupling constant that will be
determined by an appropriate variational principle.

We consider a radial foliation for the spacetime (Gaussian normal
coordinates)%

\begin{equation}
ds^{2}=N^{2}(\rho )d\rho ^{2}+h_{ij}\left( \rho ,x\right) dx^{i}dx^{j},
\label{radfol}
\end{equation}%
but we do not assume any particular expansion of the boundary
metric as in eq.(\ref{confcoord}),(\ref{FGexp}). In this frame,
the expression for the extrinsic curvature adopts a simple form%

\begin{equation}
K_{ij}=-\frac{1}{2N}\partial _{\rho }h_{ij}.
\end{equation}

\subsection{Kounterterms and the Euler Theorem}

The Kounterterms series differs from the one obtained from the
Dirichlet regularization, even in a simple case as it is $D=4$,
where it is given by the boundary term \cite{OleaJHEP}

\begin{equation}
B_{3}=2\sqrt{-h}\delta _{\lbrack
j_{1}j_{2}j_{3}]}^{[i_{1}i_{2}i_{3}]}K_{i_{1}}^{j_{1}}(R_{i_{2}i_{3}}^{j_{2}j_{3}}(h)-%
\frac{2}{3}K_{i_{2}}^{j_{2}}K_{i_{3}}^{j_{3}}),  \label{B3}
\end{equation}%
with a coupling constant $c_{3}=\ell ^{2}/(64\pi G)$. In the above formula, $%
R_{kl}^{ij}(h)$ stands for the intrinsic curvature of the boundary metric.

It is clear from the expanded form

\begin{equation}
B_{3}=4\sqrt{-h}\left[ -\frac{2}{3}%
K_{j}^{i}K_{k}^{j}K_{i}^{k}+K(K_{j}^{i}K_{i}^{j}-\frac{1}{3}%
K^{2})-2(R_{j}^{i}-\frac{1}{2}\delta _{j}^{i}R)K_{i}^{j}\right] ,
\label{B3exp}
\end{equation}%
that $B_{3}$ does not contain any term proportional to
$\sqrt{-h}K$ (Gibbons-Hawking term), making evident that it is not
derived from a
Dirichlet action principle. Notice that the dimensional continuation of (\ref%
{B3exp}) is required to define the Dirichlet problem for
Einstein-Gauss-Bonnet gravity in $D\geq 5$, because it is the
generalization of the Gibbons-Hawking term for the quadratic terms
in the curvature in the Gauss-Bonnet density \cite{Myers,MH}.

The expression (\ref{B3}) possesses the additional Lorentz
symmetry in the tangent space that becomes manifest when it is
expressed in terms of the second fundamental form (SFF)%

\begin{equation}
\theta ^{AB}=\omega ^{AB}-\bar{\omega}^{AB}, \label{SFFdef}
\end{equation}%
defined as the difference between the dynamic spin connection
$\omega ^{AB}=\omega _{\mu }^{AB}dx^{\mu }$ and a reference one
$\bar{\omega}^{AB}$.

The indices of the tangent space run in the set
$A,B=\{0,1,..,D-1\}$. In the metric formulation of gravity, the
spin connection is determined in terms of the vielbein
$e^{A}=e_{\mu }^{A}dx^{\mu }$ ($\mathcal{G}_{\mu \nu }=\eta
_{AB}e_{\mu }^{A}e_{\nu }^{B}$) as%

\begin{equation}
\omega _{\mu }^{AB}=-e^{B\nu }\nabla _{\mu }e_{\nu }^{A}.
\end{equation}%

For the radial foliation (\ref{radfol}), the natural splitting for
the orthonormal basis is

\begin{eqnarray}
e^{1} &=&Nd\rho ,  \label{e1} \\
e^{a} &=&e_{i}^{a}dx^{i},
\end{eqnarray}%
where the indices set $A=\{1,a\}$ and the boundary metric is
$h_{ij}=\eta _{ab}e_{i}^{a}e_{j}^{b}$.

An adequate choice of the reference connection
$\bar{\omega}^{AB},$ as
obtained from a cobordant product metric%

\begin{equation}
ds^{2}=\bar{N}^{2}(\rho )d\rho ^{2}+\bar{h}_{ij}(x)dx^{i}dx^{j}
\label{Ncoord}
\end{equation}%
that matches the dynamic one only on the boundary $\rho =\rho _{0}$, ($\bar{h%
}_{ij}(x)=h_{ij}(\rho _{0},x)$) leads to a SFF on the boundary as \cite%
{eguchi,Spivak,Choquet-Dewitt,Myers}

\begin{equation}
\begin{array}{cc}
\theta ^{1a}=K_{i}^{a}dx^{i}, & \theta ^{ab}=0,%
\end{array}
\label{SFFnortan}
\end{equation}%
in terms of the extrinsic curvature
$K_{i}^{a}=e_{j}^{a}K_{i}^{j}$. We
stress that the spin connection $\bar{\omega}^{AB}$ is just introduced on $%
\partial M$ to restore Lorentz covariance in the boundary term but it is not
related to any background-substraction procedure, where the
background needs to be a solution of the bulk field equations.

Therefore, the fully Lorentz-covariant expression for the Kounterterms $%
B_{3} $ in terms of the Levi-Civita tensor is \footnote{%
The wedge product $\wedge $ between the differential forms is
omitted throughout the paper.}

\begin{equation}
B_{3}=2\varepsilon _{ABCD}\theta ^{AB}(R^{CD}+\frac{1}{3}\theta
_{F}^{C}\theta ^{FD}). \label{B3CF}
\end{equation}%

In four-dimensional manifolds without boundary, the integration of
the Euler-Gauss-Bonnet term

\begin{equation}
\mathcal{E}_{4}=\varepsilon _{ABCD}\,\hat{R}%
^{AB}\hat{R}^{CD}  \nonumber \\
=-d^{4}x\sqrt{-\cal{G}}\,\left( \hat{R}%
_{\mu \nu \alpha \beta }\,\hat{R}^{\mu \nu \alpha \beta
}-4\hat{R}_{\mu \nu }\,\hat{R}^{\mu \nu }+\hat{R}^{2}\right) \,
\end{equation}%
is simply proportional to the Euler characteristic $\chi (M_{4})$.
When a boundary is introduced, the Euler theorem states that there
is a correction due to the boundary%
\begin{equation}
\int\limits_{M_{4}}\varepsilon
_{ABCD}\,\hat{R}^{AB}\hat{R}^{CD}=2\left( 4\pi \right) ^{2}\chi
(M_{4})+2\int\limits_{\partial M_{4}}\varepsilon _{ABCD}\,\theta
^{AB}\left( R^{CD}+\frac{1}{3}\,(\theta ^{2})^{CD}\right) ,
\end{equation}%
given exactly by the expression (\ref{B3CF}) that is known as the
second Chern form. In fact, using the general formalism reviewed in
Appendix \ref{KTTF}, the boundary term can be seen as a
transgression form for the Lorentz group $SO(3,1)$.

In higher even-dimensional AdS gravity, the regularization of the
conserved quantities was achieved in the Ref.\cite{ACOTZ2n} by the
addition of the Euler term $\mathcal{E}_{2n}$. This term is no
longer equivalent to the Gauss-Bonnet term, because it is the
maximal Lovelock form in that dimension. This
construction is locally equivalent, by virtue of the Euler theorem, to the $%
n-$th Chern form
\begin{equation}
B_{2n-1}=n\int_{0}^{1}dt\varepsilon _{A_{1}...A_{2n}}\theta
^{A_{1}A_{2}}(R^{A_{3}A_{4}}+t^{2}\theta _{F}^{A_{3}}\theta ^{FA_{4}})\times
...\times (R^{A_{2n-1}A_{2n}}+t^{2}\theta _{F}^{A_{2n-1}}\theta ^{FA_{2n}})
\label{B2n-1}
\end{equation}%
in terms of the continuous parameter $t$. This parametrization is
useful not only to write down a compact formula for the boundary
term but to generate the relative coefficients of the binomial
expansion, as well. This fact has a deep geometrical origin as an
explicit realization of the Cartan homotopy operator, that permits
to obtain the explicit form of a boundary term whose exterior
derivative is the difference of two invariant polynomials for a
given Lie group (See Appendix \ref{KTTF}).
 It can be shown that the above boundary term also cancels the divergences from
radial infinity in the evaluation of the bulk Euclidean action
\cite{OleaJHEP}.

\subsection{Kounterterms and Chern-Simons Forms}

Remarkably, the regularization prescription given by the maximal
Chern-form in even dimensions works equally well for
Einstein-Gauss-Bonnet gravity, where the coupling constant carried
by $B_{2n-1}$ takes a different value respect to the same problem in
Einstein-Hilbert \cite{Kof-Ole}. The same situation is found in a
generic even-dimensional Lovelock-AdS theory, where the Kounterterms
series (\ref{B2n-1}) preserves its form, but again its coupling
constant changes accordingly \cite{KO-Lovelock}. This fact strongly
suggests the universality of the boundary terms that regularize the
action for a set of inequivalent gravity theories (at least, the
ones that are Lovelock-type).

On the other hand, a finite action principle was set for
Chern-Simons-AdS gravity, a particular Lovelock theory in
odd-dimensions that possesses a unique cosmological constant,
contains higher powers in the curvature and can be obtained from a
Chern-Simons form for the AdS group connection. The guiding line to
derive the correct form of the boundary terms is restoring gauge
invariance by the use of transgression forms.

As we can see from Appendix \ref{KTTF}, the expression for the
Kounterterms in Chern-Simons-AdS gravity is given by a double
integral
in the parameters $t,s\in \lbrack 0,1]$%
\begin{eqnarray}
B_{2n} &=&n\int_{0}^{1}dt\int_{0}^{t}ds\varepsilon _{A_{1}...A_{2n+1}}\theta
^{A_{1}A_{2}}e^{A_{3}}(R^{A_{4}A_{5}}+t^{2}\theta _{F}^{A_{4}}\theta
^{FA_{5}}+\frac{s^{2}}{\ell ^{2}}e^{A_{4}}e^{A_{5}})\times ...  \notag \\
&&...\times (R^{A_{2n}A_{2n+1}}+t^{2}\theta _{F}^{A_{2n}}\theta ^{FA_{2n+1}}+%
\frac{s^{2}}{\ell ^{2}}e^{A_{2n}}e^{A_{2n+1}}).  \label{B2ntheta}
\end{eqnarray}%
In the spirit of the alternative regularization of AdS gravity in
even dimensions, we will assume universality of the form of the
regularizing boundary terms in $D=2n+1$ dimensions. In fact, as we
shall explicitly demonstrate below, the boundary term
(\ref{B2ntheta}) also leads to a finite, well-defined action
principle in odd-dimensional Einstein-Hilbert for a suitable choice
of its coupling constant $c_{2n}$.

The relations (\ref{SFFnortan}) leave a residual Lorentz symmetry on $%
\partial M$ and then, the boundary term can also be written as%
\begin{eqnarray}
B_{2n} &=&-2n\int_{0}^{1}dt\int_{0}^{t}ds\varepsilon
_{a_{1}...a_{2n}}K^{a_{1}}e^{a_{2}}(R^{a_{3}a_{4}}-t^{2}K^{a_{3}}K^{a_{4}}+%
\frac{s^{2}}{\ell ^{2}}e^{a_{3}}e^{a_{4}})\times ...  \notag \\
&&...\times (R^{a_{2n-1}a_{2n}}-t^{2}K^{a_{2n-1}}K^{a_{2n}}+\frac{s^{2}}{%
\ell ^{2}}e^{a_{2n-1}}e^{a_{2n}}),  \label{B2ninter}
\end{eqnarray}%
where $R^{ab}$ is the boundary 2-form curvature, related to the
intrinsic curvature by
$R^{ab}=\frac{1}{2}R_{ij}^{kl}e_{k}^{a}e_{l}^{b}dx^{i}\wedge
dx^{j}$.

The tensorial form of the Kounterterms can be worked out
projecting all the quantities in the boundary indices (see
Appendix \ref{UI})
\begin{eqnarray}
B_{2n} &=&2n\sqrt{-h}\int_{0}^{1}dt\int_{0}^{t}ds\delta _{\lbrack
j_{1}...j_{2n-1}]}^{[i_{1}...i_{2n-1}]}K_{i_{1}}^{j_{1}}(\frac{1}{2}%
R_{i_{2}i_{3}}^{j_{2}j_{3}}-t^{2}K_{i_{2}}^{j_{2}}K_{i_{3}}^{j_{3}}+\frac{%
s^{2}}{\ell ^{2}}\delta _{i_{2}}^{j_{2}}\delta _{i_{3}}^{j_{3}})\times ...
\notag \\
&&...\times (\frac{1}{2}%
R_{i_{2n-2}i_{2n-1}}^{j_{2n-2}j_{2n-1}}-t^{2}K_{i_{2n-2}}^{j_{2n-2}}K_{i_{2n-1}}^{j_{2n-1}}+%
\frac{s^{2}}{\ell ^{2}}\delta _{i_{2n-2}}^{j_{2n-2}}\delta
_{i_{2n-1}}^{j_{2n-1}}),  \label{B2ntensor}
\end{eqnarray}

In the language of differential forms, the gravitational action
(\ref{IgD}) can be written in terms of the local orthonormal frame
$e^{A}=e_{\mu
}^{A}dx^{\mu }$ and the 2-form Lorentz curvature $\hat{R}^{AB}=\frac{1}{2}%
\hat{R}_{\mu \nu }^{AB}dx^{\mu }\wedge dx^{\nu }$ (constructed up
from the spin connection as $\hat{R}^{AB}=d\omega ^{AB}+\omega
_{C}^{A}\omega ^{CB}$) as%

\begin{equation}
I=\kappa _{D}\int_{M}\varepsilon _{A_{1}...A_{D}}\left( \hat{R}^{A_{1}A_{2}}+%
\frac{\left( D-2\right) }{D\ell ^{2}}e^{A_{1}}e^{A_{2}}\right)
e^{A_{3}}...e^{A_{D}}+c_{d}\int_{\partial M}B_{d}.  \label{EHD}
\end{equation}%
where the constant $\kappa _{D}=(16\pi G(D-2)!)^{-1}$ and with the
boundary term $B_{d}$ given by (\ref{B2n-1}) and (\ref{B2ntheta})
for even and odd dimensions, respectively. The Lorentz curvature
is related to the spacetime Riemann tensor by
$\hat{R}^{AB}=\frac{1}{2}\hat{R}_{\mu \nu }^{\alpha \beta
}e_{\alpha }^{A}e_{\beta }^{B}dx^{\mu }\wedge dx^{\nu }$.

An arbitrary variation of the above action produces the Einstein
equations plus a surface term $\Theta $
\begin{equation}
\delta I=\int_{M}E_{A}\delta e^{A}+d\Theta ,  \label{varEHAdS}
\end{equation}%

where $E_{A}\delta e^{A}$ is the Einstein equation,

\begin{eqnarray}
E_{A}\delta e^{A} &=&\frac{1}{16\pi G(D-3)!}\varepsilon
_{AA_{2}...A_{D}}\delta e^{A}\left( \hat{R}^{A_{2}A_{3}}+\frac{1}{\ell ^{2}}%
e^{A_{2}}e^{A_{3}}\right) e^{A_{4}}...e^{A_{D}},  \label{ea} \\
&=&\frac{1}{16\pi G}\sqrt{-\mathcal{G}}\left( \hat{R}_{\mu \nu }-\frac{1}{2}%
\hat{R}\mathcal{G}_{\mu \nu }-\Lambda \mathcal{G}_{\mu \nu
}\right) \delta \mathcal{G}^{\mu \nu }.  \label{EinsEq}
\end{eqnarray}

In the Palatini formulation of gravity, the contribution to $\Theta $ coming
from the bulk term is obtained from the variation of the Riemann tensor $%
\delta \hat{R}_{\,\beta \mu \nu }^{\alpha }=\nabla _{\mu }\delta \hat{\Gamma}%
_{\beta \nu }^{\alpha }-\nabla _{\nu }\delta \hat{\Gamma}_{\beta
\mu }^{\alpha }$. For the radial foliation (\ref{radfol}), the
surface term will involve only certain components of the
connection $\hat{\Gamma}_{\mu \nu
}^{\alpha }$, related to the extrinsic curvature as%

\begin{equation}
\,\hat{\Gamma}_{ij}^{\rho }=\frac{1}{N}K_{ij},\qquad
\hat{\Gamma}_{\rho j}^{i}=-NK_{j}^{i},
\end{equation}%
such that it can be written as

\begin{equation}
\Theta =-\varepsilon _{a_{1}a_{2}...a_{d}}\delta
K^{a_{1}}e^{a_{2}}\ldots e^{a_{d}}+c_{d}\delta B_{d}.
\label{thetab}
\end{equation}

The same coordinates frame implies that the components of the
Lorentz curvature $\hat{R}^{AB}$ projected on $\partial M$ are

\begin{eqnarray}
\hat{R}^{1a} &=&DK^{a}=D_{i}K_{j}^{a}dx^{i}\wedge dx^{j},  \label{Cod} \\
\hat{R}^{ab} &=&R^{ab}-K^{a}K^{b}=\left( \frac{1}{2}%
R_{ij}^{ab}-K_{i}^{a}K_{j}^{b}\right) dx^{i}\wedge dx^{j}
\label{GaCod}
\end{eqnarray}%
where $D_{i}$ is the covariant derivative in the boundary indices
(defined with the submanifold spin connection $\omega ^{ab}$) and we
have dropped the components along $d\rho $. Eqs.(\ref{Cod}) and
(\ref{GaCod}) are just the
well-known Gauss-Coddazzi relations for a radial foliation (\ref{radfol})%

\begin{eqnarray}
\hat{R}_{ij}^{\rho l} &=&-\frac{1}{N}\nabla _{\lbrack
i}K_{j]}^{l},
\label{Coddazzi} \\
\hat{R}_{ij}^{kl}
&=&R_{ij}^{kl}-K_{i}^{k}K_{j}^{l}+K_{i}^{l}K_{j}^{k}.
\label{GCtensor}
\end{eqnarray}%

As an illustrative, simple example of the present procedure, in
the next section we show explicitly how the Kounterterms series in
five dimensions leads to a well-posed action principle for
boundary conditions derived from the asymptotic form for AAdS
spacetimes (\ref{confcoord}),(\ref{FGexp}).

\section{Five-Dimensional Case}

Dirichlet counterterms are local functional that preserve general
covariance at the boundary. Kounterterms respect
Lorentz-covariance in the tangent
space, what provides a criterion to select them. From the surface term (\ref%
{thetab}), it is clear that $B_{d}$ must be constructed with the
same parity as the bulk action, that is, the same invariant tensor
$\epsilon _{A_{1}...A_{D}}$ for the Lorentz group (and not $\delta
_{\lbrack CD]}^{[AB]}$). In particular, this argument rules out
the addition of topological invariants of the Pontryagin class in
even dimensions and many possible boundary terms in the case we
are treating here.

Kounterterms are built up as totally antisymmetric $2n-$forms.
This eliminates a possible inclusion of terms containing covariant
derivatives of the intrinsic curvature, which are discarded by the
Bianchi identity $\nabla_{\lbrack m}R_{ij]}^{kl}=0$.

The extrinsic curvature can be defined in an arbitrary frame as $%
K_{AB}=-h_{A}^{C}h_{B}^{D}n_{C;D}$, where $n^{A}$ is a unit vector
normal to
the boundary, and related to the SFF by $\theta ^{AB}=n^{A}K^{B}-n^{B}K^{A}$%
, ($K^{A}=K_{B}^{A}e^{B}$). In that way, we can always write down
the Kounterterms as a fully-covariant expression, independent of
any particular
foliation. In five dimensions this is given by%

\begin{equation}
B_{4}=\epsilon _{A_{1}...A_{5}}\theta ^{A_{1}A_{2}}e^{A_{3}}\Big(%
R^{A_{4}A_{5}}+\frac{1}{2}\theta _{\,\,\,\,C}^{A_{4}}\theta ^{CA_{5}}+\frac{1%
}{6\ell ^{2}}e^{A_{4}}e^{A_{5}}\Big),
\end{equation}%
with the equivalence in tensorial notation (see Appendix \ref{UI})%

\begin{equation}
B_{4}=\sqrt{-h}\delta _{\lbrack
j_{1}j_{2}j_{3}]}^{[i_{1}i_{2}i_{3}]}K_{i_{1}}^{j_{1}}\Big(%
R_{i_{2}i_{3}}^{j_{2}j_{3}}-K_{i_{2}}^{j_{2}}K_{i_{3}}^{j_{3}}+\frac{1}{%
3\ell ^{2}}\delta _{i_{2}}^{j_{2}}\delta _{i_{3}}^{j_{3}}\Big).
\label{B4}
\end{equation}

\subsection{Variational principle and asymptotic conditions}

Arbitrary variations of the total action (\ref{IgD}) produce a surface term%

\begin{eqnarray}
\delta I_{5} &=&-2\int_{\partial M}\kappa _{_{5}}\varepsilon
_{abcd}\delta
K^{a}e^{b}e^{c}e^{d}+c_{4}\epsilon _{abcd}\delta K^{a}e^{b}\Big(R^{cd}-\frac{%
3}{2}K^{c}K^{d}+\frac{1}{6\ell ^{2}}e^{c}e^{d}\Big)  \notag \\
&&+c_{4}\varepsilon _{abcd}K^{a}\delta e^{b}\Big(R^{cd}-\frac{1}{2}%
K^{c}K^{d}+\frac{1}{2\ell ^{2}}e^{c}e^{d}\Big),  \label{var5}
\end{eqnarray}%
when equations of motion hold. The constant in front of the bulk action is $%
\kappa _{5}=1/(96\pi G)$. The above equation can be conveniently rewritten as%
\begin{eqnarray}
\delta I_{5} &=&-2\int_{\partial M}\kappa _{_{5}}\varepsilon
_{abcd}\delta
K^{a}e^{b}e^{c}e^{d}+2c_{4}\epsilon _{abcd}\delta K^{a}e^{b}\Big(\hat{R}%
^{cd}+\frac{1}{3\ell ^{2}}e^{c}e^{d}\Big) \\
&&-c_{4}\varepsilon _{abcd}(\delta K^{a}e^{b}-K^{a}\delta e^{b})\Big(R^{cd}-%
\frac{1}{2}K^{c}K^{d}+\frac{1}{2\ell ^{2}}e^{c}e^{d}\Big),
\label{prevar5}
\end{eqnarray}%
with the help of the Gauss-Coddazzi relation (\ref{GaCod}).

A well-posed variational principle for precise asymptotic
conditions in AdS gravity is essential to attain the finiteness of
the conserved quantities and the Euclidean action. This amounts to
on-shell cancelation of the surface term (\ref{var5}) using
boundary conditions derived from the asymptotic form of AAdS
spacetimes (\ref{confcoord}),(\ref{FGexp}).

In general, the Dirichlet variational problem for gravity is
well-defined if one supplements the action by the Gibbons-Hawking
term and fixes the metric $h_{ij}$ at the boundary. In this way, the
surface term coming from an arbitrary variation of the Dirichlet
action vanishes identically, no matter if the boundary is at a
finite distance (e.g., on a brane, to define the Israel matching
conditions) or at infinity. However, it has been recently argued in
ref.\cite{Pa-Sk1} that the standard Dirichlet boundary condition for
the metric $h_{ij}$ does not really make sense for manifolds with
conformal boundary, as it is the case of AAdS spaces. It is clear
from eqs.(\ref{confcoord}),(\ref{FGexp}) that the variation of
$h_{ij}$ is divergent at $\rho=0$ and one should instead fix the
conformal structure $g_{(0)ij}$. However, due to the divergence
produced in this way, the action requires additional boundary terms
on top of the Gibbons-Hawking term, that turn out to be the standard
Dirichlet counterterms.

In the present formulation, we will consider a boundary condition
that derives from the asymptotic expansion of the extrinsic curvature%

\begin{equation}
K_{j}^{i}=K_{jl}h^{li}=\frac{1}{\ell }\delta _{j}^{i}-\frac{\rho }{\ell }%
(g_{(0)}^{-1}g_{(1)})_{j}^{i}-\frac{\rho ^{2}}{\ell }%
(2g_{(0)}^{-1}g_{(2)}-g_{(0)}^{-1}g_{(1)}g_{(0)}^{-1}g_{(1)})_{j}^{i}+...,
\label{kfg}
\end{equation}%
in increasing powers of $\rho $. This asymptotic behavior implies
that the
extrinsic curvature satisfies%

\begin{equation}
K_{i}^{j}=\frac{1}{\ell }\delta _{i}^{j},  \label{Kdelta}
\end{equation}%
on the conformal boundary, and an arbitrary variation is given by

\begin{equation}
\delta K_{i}^{j}=0.  \label{delKdelta}
\end{equation}
Had we attempted to fix
$K_{ij}=\frac{1}{\ell}\frac{g_{(0)ij}}{\rho}+...$, we would have
faced the same problem as fixing the boundary metric $h_{ij}$. On
the contrary, eq.(\ref{delKdelta}) is a regular boundary condition
on the extrinsic curvature that can be derived from fixing
$g_{(0)ij}$ due to the asymptotic form of AAdS spaces. In
particular, in the asymptotically flat limit ($\ell \rightarrow
\infty$) this accident no longer occurs.

 The condition (\ref{Kdelta}) has been also taken as
the boundary data for the problem of holographic reconstruction of
the spacetime in terms of the extrinsic curvature in
ref.\cite{Pa-Sk1}.

Making explicit the indices in the term proportional to the \textit{curl} $%
\varepsilon _{abcd}(\delta K^{a}e^{b}-K^{a}\delta e^{b})$, the
last line of
eq.(\ref{prevar5}) is%

\begin{equation}
\varepsilon _{abcd}\,\varepsilon ^{i_{1}i_{2}i_{3}i_{4}}\left[
\delta K_{i_{1}}^{j}e_{j}^{a}e_{i_{2}}^{b}+\delta
e_{j}^{a}e_{l}^{b}\left(
K_{i_{1}}^{j}\delta _{i_{2}}^{l}-K_{i_{2}}^{l}\delta _{i_{1}}^{j}\right) %
\right] \left(
R_{i_{3}i_{4}}^{cd}-K_{i_{3}}^{c}K_{i_{4}}^{d}+\frac{1}{\ell
^{2}}\,e_{i_{3}}^{c}e_{i_{4}}^{d}\right) d^{4}x,
\end{equation}%
that vanishes identically when we take the condition
(\ref{Kdelta}) on the extrinsic curvature and its variation
(\ref{delKdelta}). We assume a constant (negative) curvature in
the asymptotic region (\ref{Reqdelta}),
that in the language of differential forms reads%

\begin{equation}
\hat{R}^{AB}+\frac{1}{\ell ^{2}}e^{A}e^{B}=0,  \label{ALAdSdf}
\end{equation}%
that in particular holds for the boundary indices. This is a local
condition at the boundary known as \textbf{ALAdS} (asymptotically
locally AdS) that in principle does not impose further
restrictions on the global topology of the spacetime. Therefore,
solutions of this class include not only point-like black holes as
Schwarzschild-AdS and Kerr-AdS, but also extended objects as black
strings.

The coupling constant $c_{4}$ is then fixed as

\begin{equation}
c_{4}=\frac{3\kappa _{5}\ell ^{2}}{4}=\frac{\ell ^{2}}{128\pi G},
\end{equation}%
to cancel the rest of the surface term.

Now that we know the coefficient in front of the boundary term, we
notice that the term proportional to $\sqrt{-h}K$ carries an
\textit{anomalous} factor $\frac{1}{64\pi G}$ compared to the one of
the Gibbons-Hawking term in eq.(\ref{Ict}), what is a consequence of
a different action principle.

\subsection{Conserved quantities}

In the standard Dirichlet problem of gravity, the bulk contribution to (\ref%
{prevar5}) is canceled by the Gibbons-Hawking term and no other
terms along $\delta K^{a}$ can appear from the boundary term
$\mathcal{B}_{ct}$, as it is a functional only of intrinsic
quantities.

In the present approach, the surface term (\ref{prevar5}) contains
variations along the extrinsic curvature, such that we cannot
identify a quasilocal (boundary) stress tensor from the variation
of the total action.

However, we can always define the energy and other conserved
charges associated to asymptotic symmetries of a gravitational
system through the Noether theorem.

In the Appendix \ref{Noether} we summarize the construction of the
Noether charges for an arbitrary Lagrangian. The conserved current
in the case we are
considering here is given by%

\begin{equation}
\ast J=-\Theta (e^{a},K^{a},\delta e^{a},\delta K^{a})-i_{\xi
}(L_{5}+c_{4}dB_{4}),
\end{equation}%
where $L_{5}$ is the bulk Lagrangian, $i_{\xi }$ is the interior
derivative (also known as contraction operator) with the Killing
vector $\xi ^{\mu }$
defined in the Appendix \ref{Noether} and $\Theta $ is the surface term%

\begin{eqnarray}
\Theta &=&\frac{\ell ^{2}}{32\pi G}\left[ \epsilon _{abcd}\delta K^{a}e^{b}%
\Big(\hat{R}^{cd}+\frac{1}{\ell ^{2}}e^{c}e^{d}\Big)\right.  \notag \\
&&\left. -\frac{1}{2}\epsilon _{abcd}(\delta K^{a}e^{b}-K^{a}\delta e^{b})%
\Big(R^{cd}-\frac{1}{2}K^{c}K^{d}+\frac{1}{2\ell
^{2}}e^{c}e^{d}\Big)\right] .  \label{surt5}
\end{eqnarray}%

The derivation of the conserved charges from the current extensively
uses the properties of interior, exterior and Lie derivatives,
Bianchi identity, and the equations of motion in differential forms
language. However, we will exploit a shortcut for the charges
(\ref{Qnew}) pointed out in Appendix \ref{Noether}, by identifying
the contributions coming from the bulk Lagrangian and the boundary
term $B_{4}$.

In doing so, the Noether charge is written as%

\begin{equation}
Q(\xi )=\mathcal{K}(\xi )+c_{4}\int_{\partial \Sigma }\left( i_{\xi }K^{a}%
\frac{\delta B_{4}}{\delta K^{a}}+i_{\xi }e^{a}\frac{\delta
B_{4}}{\delta e^{a}}\right)  \label{chargegeneral}
\end{equation}%
where the first term is known as the Komar's integral%

\begin{equation}
\mathcal{K}(\xi )=\frac{1}{48\pi G}\int_{\partial \Sigma
}\varepsilon _{abcd}\,i_{\xi }K^{a}e^{b}e^{c}e^{d},
\end{equation}%
and it is the conserved quantity associated to the bulk term in
the gravity
action.\footnote{%
In an arbitrary dimension $D>3$, the Komar's integral has the form
\begin{equation*}
\mathcal{K}(\xi )=\frac{1}{8\pi G_{N}}\int_{\partial \Sigma
}\varepsilon _{a_{1}...a_{D-1}}\,i_{\xi
}K^{a_{1}}e^{a_{2}}...e^{a_{D-1}}=\frac{1}{16\pi
G_{N}}\int_{\partial \Sigma }\nabla ^{\mu }\xi ^{\nu }d\Sigma
_{\mu \nu }\,,
\end{equation*}%
where $d\Sigma _{\mu \nu }=\frac{\sqrt{\sigma }}{(D-2)!}\epsilon
_{\mu \nu \alpha _{1}...\alpha _{D-2}}dx^{\alpha _{1}}\wedge
...\wedge dx^{\alpha _{D-2}}$ is the dual of the area $(D-2)$-form
($\sigma $ is the determinant of the metric on the sphere
$S^{D-2}$). When evaluated for a timelike Killing vector on the
Schwarzschild-AdS solution, this formula gives a factor
$(D-3)/(D-2)$ times the mass, plus a divergence in the radial
coordinate. The divergence is usually canceled by background
substraction, procedure that however does not solve the problem of
the anomalous Komar factor \cite{K-KBL}.}

The expression for the Noether charge appears to be naturally
split in two parts, associated to the first and second line of the
surface term (\ref{surt5}), respectively
\begin{equation}
Q(\xi )=q(\xi )+q_{0}(\xi ),
\end{equation}%

\begin{eqnarray}
q(\xi ) &=&\frac{\ell ^{2}}{32\pi G}\int_{\partial \Sigma
}\epsilon
_{abcd}i_{\xi }K^{a}e^{b}\Big(\hat{R}^{cd}+\frac{1}{\ell ^{2}}e^{c}e^{d}\Big)%
,  \label{qxi5} \\
q_{0}(\xi ) &=&-\frac{\ell ^{2}}{64\pi G}\int_{\partial \Sigma
}\epsilon
_{abcd}\left( i_{\xi }K^{a}e^{b}+K^{a}i_{\xi }e^{b}\right) \Big(R^{cd}-\frac{%
1}{2}K^{c}K^{d}+\frac{1}{2\ell ^{2}}e^{c}e^{d}\Big).
\label{q0xi5}
\end{eqnarray}%

As we shall see below for concrete examples, the formula for
$q(\xi )$ provides the standard conserved quantities (mass,
angular momentum) for AAdS solutions. In tensorial notation,
$q(\xi )$ takes the form

\begin{equation}
q(\xi )=\frac{\ell ^{2}}{64\pi G}\int_{\partial \Sigma
}\sqrt{-h}\epsilon
_{i_{1}...i_{4}}(\xi ^{k}K_{k}^{i_{1}})\delta _{j_{2}}^{i_{2}}\Big(\hat{R}%
_{j_{3}j_{4}}^{i_{3}i_{4}}+\frac{1}{\ell ^{2}}\delta _{\lbrack
j_{3}j_{4}]}^{[i_{3}i_{4}]}\Big)dx^{j_{2}}dx^{j_{3}}dx^{j_{4}},
\label{qxitensor5}
\end{equation}%
where the product $dx^{j_{2}}\wedge dx^{j_{3}}\wedge dx^{j_{4}}$
stands for the volume element of the boundary of the spatial
section $\partial \Sigma $ (i.e., boundary indices without the
time). Notice that (\ref{qxitensor5}) is proportional to the
l.h.s. of eq.(\ref{Reqdelta}), and therefore vanishes identically
for any solution of global constant curvature, as AdS vacuum and
spacetimes with topological identifications that preserve AdS
flatness.

The above reasoning leads us to consider (\ref{q0xi5}) as a
general formula for the vacuum energy in five-dimensional AAdS
spacetimes

\begin{equation}
q_{0}(\xi )=-\frac{\ell ^{2}}{128\pi G}\int_{\partial \Sigma }\sqrt{-h}%
\epsilon _{i_{1}i_{2}i_{3}i_{4}}\xi ^{k}(K_{k}^{i_{1}}\delta
_{j_{2}}^{i_{2}}+K_{j_{2}}^{i_{1}}\delta _{k}^{i_{2}})\Big(%
R_{j_{3}j_{4}}^{i_{3}i_{4}}-\frac{1}{2}K_{[j_{3}j_{4}]}^{[i_{3}i_{4}]}+\frac{%
1}{2\ell ^{2}}\delta _{\lbrack j_{3}j_{4}]}^{[i_{3}i_{4}]}\Big)%
dx^{j_{2}}dx^{j_{3}}dx^{j_{4}},  \label{q0xitensor5}
\end{equation}%
where we have introduced the shorthand%

\begin{equation}
K_{[kl]}^{[ij]}=K_{k}^{i}K_{l}^{j}-K_{k}^{j}K_{l}^{i}.
\end{equation}%

The regularization of the conserved quantities and the Euclidean
action in five-dimensional AAdS solutions is illustrated through
concrete examples below.

\subsection{Examples}

-\textit{Schwarzschild-AdS black hole and topological extensions}

Static solutions to EH-AdS gravity are given by Schwarzschild
black hole
metric. Due to the negative cosmological constant, the transversal section $%
\Sigma _{D-2}^{k}$ can be the sphere $S^{D-2}$, a $(D-2)$-dim
locally flat space or the hyperboloid $H^{D-2}$%

\begin{equation}
ds^{2}=-\Delta ^{2}(r)dt^{2}+\frac{dr^{2}}{\Delta ^{2}(r)}+r^{2}\gamma _{%
\underline{m}\underline{n}}d\theta ^{\underline{m}}d\theta
^{\underline{n}}, \label{Sch-AdS}
\end{equation}%
where the metric function is

\begin{equation}
\Delta ^{2}(r)=k-\frac{2\omega _{D}G\mu
}{r^{D-3}}+\frac{r^{2}}{\ell ^{2}}. \label{Delta2}
\end{equation}%

Here, $\mu $ appears as an integration constant, $\omega _{D}=\frac{8\pi }{%
(D-2)Vol(S^{D-2})}$ and $\gamma _{\underline{m}\,\underline{n}}$ ($%
\underline{m},\underline{n}=1,...,D-2$) is the metric of $\Sigma
_{D-2}^{k}$ of constant curvature $k=\pm 1,0$. The event horizon
$r_{+}$ is defined as the largest root of $\Delta ^{2}(r_{+})=0$.

In order to compute the mass and the vacuum energy, we use the following
components of the extrinsic and intrinsic curvatures%

\begin{equation}
K_{t}^{t}=-\Delta ^{\prime },\qquad K_{\underline{m}}^{\underline{n%
}}=-\frac{\Delta }{r}\delta _{\underline{m}}^{\underline{n}},
\label{KSch}
\end{equation}%

\begin{equation}
R_{\underline{m}_{2}\underline{n}_{2}}^{\underline{m}_{1}\underline{n}_{1}}=%
\frac{k}{r^{2}}\delta _{\lbrack \underline{m}_{2}\underline{n}_{2]}}^{[%
\underline{m}_{1}\underline{n}_{1]}}.  \label{Rsphere}
\end{equation}%
where prime denotes the derivative $d/dr$, and the determinant of
the
boundary metric%

\begin{equation}
\sqrt{-h}=\Delta \sqrt{\gamma }r^{D-2}.  \label{sqrth}
\end{equation}

Notice the opposite sign in the leading order of the asymptotic behavior of $%
K_{j}^{i}$ respect to eq.(\ref{kfg}), due to the change in the
radial coordinate.

Thus, for the time-like Killing vector $\xi =\partial /\partial t$,
formula (\ref{qxitensor5}) gives%

\begin{eqnarray}
q(\partial _{t}) &=&M=3!Vol(\Sigma _{3}^{k})\lim_{r\rightarrow \infty
}(\Delta ^{2})^{\prime }r\Big\{\kappa _{_{5}}r^{2}+2c_{4}(k-\Delta ^{2}+%
\frac{r^{2}}{3\ell ^{2}})\Big\},  \label{Massint5} \\
&=&\frac{Vol(\Sigma _{3}^{k})}{Vol(S^{3})}\mu ,  \label{Mass5}
\end{eqnarray}%
in agreement with the result from background-dependent methods,
e.g., Hamiltonian formalism \cite{H-TAdS}. We have written
eq.(\ref{Massint5}) as an intermediate step just to make clear the
contributions from the bulk and the boundary that will be useful
for the black hole thermodynamics below.

Plugging the metric (\ref{Sch-AdS}) into the formula
(\ref{q0xitensor5}), we obtain%

\begin{eqnarray}
q_{0}(\partial _{t}) &=&E_{0}=-3!Vol(\Sigma _{3}^{k})c_{4}\lim_{r\rightarrow
\infty }\Big(\Delta ^{2}-\frac{r(\Delta ^{2})^{\prime }}{2}\Big)\Big(k-\frac{%
1}{2}\Delta ^{2}+\frac{r^{2}}{2\ell ^{2}}\Big)  \label{e05} \\
&=&(-k)^{2}\frac{3Vol(\Sigma _{3}^{k})}{64\pi G}\ell ^{2}.  \label{vac5}
\end{eqnarray}%

The result for the spherical case ($k=1$) is the
Balasubramanian-Kraus vacuum energy $E_{0}=\frac{3\pi \ell
^{2}}{32G}$ for SAdS black hole that appears in the Dirichlet
regularization of the stress tensor \cite{Ba-Kr}.

In the case $k=-1$, cosmic censorship holds for a mass over a
critical value $M_{c}=\frac{Vol(H^{3})}{Vol(S^{3})}\mu _{c}$ that
separates black holes with hyperbolic transversal section from
naked singularities. This value, for an arbitrary dimension, is%

\begin{equation}
\mu _{c}=-\frac{\ell ^{D-3}}{\omega _{D}G}\sqrt{\frac{(D-3)^{D-3}}{%
(D-1)^{D-1}},}  \label{muc}
\end{equation}%
where $\omega _{D}=\frac{8\pi }{(D-2)Vol(S^{D-2})}$. The critical
mass in five dimensions has the same value (with opposite sign) as
the vacuum energy (\ref{vac5}). Therefore, despite the fact that
in $5D$ AdS gravity there exist black holes with negative mass,
the vacuum energy restores the positivity of the total energy
$E=M+E_{0}$ \cite{Emparan-Johnson-Myers}.

We show now that the boundary term (\ref{B4}) makes finite the
Euclidean action (\ref{IgD}) for five-dimensional SAdS solution,
and that correctly describes black hole thermodynamics.

The Euclidean period $\beta $ is defined as%

\begin{equation}
\beta =T^{-1}=\frac{4\pi }{(\Delta ^{2})^{\prime }\Big|_{r_{+}}}
\label{betaSAdS}
\end{equation}%
where $T$ is the black hole temperature. This condition comes
from the requirement that, in the Euclidean sector the solution (\ref%
{Sch-AdS}) does not have a conical singularity at the coordinates origin ($%
r=r_{+}$). In the canonical ensemble, the Euclidean action

\begin{equation}
S=\beta \mathcal{E}-I^{E},  \label{SIE}
\end{equation}%
defines the entropy $S$ and the thermodynamic energy

\begin{equation}
\mathcal{E}=-\frac{\partial I^{E}}{\partial \beta }
\label{thermoE}
\end{equation}%
of a black hole for a fixed temperature. The Euclidean bulk action
is evaluated for a static black hole of the form (\ref{Sch-AdS})
as a total derivative in the radial coordinate, such that

\begin{equation}
I_{bulk}^{E}=-\kappa _{_{5}}3!Vol(\Sigma _{3}^{k})\beta \left.
\lbrack (\Delta ^{2})^{\prime }r^{3}]\right\vert _{r_{+}}^{\infty
},  \label{IEbulk5}
\end{equation}%
and the Euclidean boundary term as

\begin{equation}
\int_{\partial {M}}B_{4}^{E}=-3!Vol(\Sigma _{3}^{k})\beta
\Big[r(\Delta ^{2})^{\prime }\Big(k-\Delta ^{2}+\frac{r^{2}}{3\ell
^{2}}\Big)+\Big(\Delta
^{2}-\frac{r(\Delta ^{2})^{\prime }}{2}\Big)\Big(k-\frac{1}{2}\Delta ^{2}+%
\frac{r^{2}}{2\ell ^{2}}\Big)\Big]\Big|^{\infty }.  \label{BE4}
\end{equation}%

The total Euclidean action
$I_{5}^{E}=I_{bulk}^{E}+c_{4}\int_{\partial M}B_{4}^{E}$ contains
two contributions: the first one from the bulk at
radial infinity plus the boundary term that can be identified (using eqs. (%
\ref{Massint5}) and (\ref{e05}) as $-\beta (M\!+\!E_{0})$.

Therefore, the finiteness of the Noether charges for static black
holes ensures that the divergencies at $r=\infty $ of the bulk
Euclidean action are exactly canceled by the ones in the boundary
term $B_{4}$.

It is reassuring to check that the thermodynamic energy definition

\begin{equation}
\mathcal{E}=-\frac{\partial I^{E}/\partial r_{+}}{\partial \beta
/\partial r_{+}}=M+E_{0},  \label{thermoenergy}
\end{equation}%
recovers the same result for the total energy as from the Noether
charges defined above.

Finally, the entropy is the horizon contribution of the bulk
Euclidean action

\begin{equation}
S=\frac{vol(\Sigma _{3}^{k})\,r_{+}^{3}}{4G}=\frac{Area}{4G}.
\label{entro5}
\end{equation}

\textit{-Kerr-AdS black hole}

The general Kerr-AdS metric in five dimensions possesses two
rotation parameters $a,b$, and it can be written in
Boyer-Lindquist coordinates as \cite{HHTR}

\begin{eqnarray}
ds^{2} &=&-\frac{\Delta _{r}}{\rho ^{2}}\left( dt-\frac{a\sin ^{2}\theta }{%
\Xi _{a}}d\phi -\frac{b\cos ^{2}\theta }{\Xi _{b}}d\psi \right) ^{2}+\frac{%
\rho ^{2}dr^{2}}{\Delta _{r}}+\frac{\rho ^{2}d\theta ^{2}}{\Delta _{\theta }}%
+  \notag \\
&&+\frac{\Delta _{\theta }\sin ^{2}\theta }{\rho ^{2}}\left( adt-\frac{%
\left( r^{2}+a^{2}\right) }{\Xi _{a}}d\phi \right)
^{2}+\frac{\Delta _{\theta }\cos ^{2}\theta }{\rho ^{2}}\left(
bdt-\frac{\left(
r^{2}+b^{2}\right) }{\Xi _{b}}d\psi \right) ^{2}+  \notag \\
&&+\frac{\left( 1+r^{2}/\ell ^{2}\right) }{r^{2}\rho ^{2}}\left( abdt-\frac{%
b\left( r^{2}+a^{2}\right) \sin ^{2}\theta }{\Xi _{a}}d\phi
-\frac{a\left( r^{2}+b^{2}\right) \cos ^{2}\theta }{\Xi _{b}}d\psi
\right) ^{2}, \label{KerrAdS5}
\end{eqnarray}%
where the functions in the metric are%

\begin{eqnarray}
\Delta _{r} &\equiv &\frac{1}{r^{2}}\left( r^{2}+a^{2}\right)
\left(
r^{2}+b^{2}\right) \left( 1+r^{2}/\ell ^{2}\right) -2m,  \label{Deltar} \\
\Delta _{\theta } &\equiv &1-\frac{a^{2}}{\ell ^{2}}\cos ^{2}\theta -\frac{%
b^{2}}{\ell ^{2}}\sin ^{2}\theta ,  \label{DeltaTheta} \\
\rho ^{2} &\equiv &r^{2}+a^{2}\cos ^{2}\theta +b^{2}\sin
^{2}\theta ,
\label{rhodef} \\
\Xi _{a} &\equiv &1-\frac{a^{2}}{\ell ^{2}},\qquad \Xi _{b}\equiv 1-\frac{%
b^{2}}{\ell ^{2}},  \label{bladef} \\
0 &\leq &\theta \leq \pi /2,\qquad 0\leq \phi \leq 2\pi ,\qquad
0\leq \psi \leq 2\pi .  \label{rangeKerr5}
\end{eqnarray}%

The event horizon $r_{+}$ is the largest solution of the equation $\Delta
_{r}(r_{+})=0$, whose area is

\begin{equation}
Area=\frac{2\pi ^{2}(r_{+}^{2}+a^{2})(r_{+}^{2}+b^{2})}{r_{+}\Xi _{a}\Xi _{b}%
}.  \label{areaKerr5}
\end{equation}%

Evaluating the charge formula (\ref{qxitensor5}) for the metric (\ref%
{KerrAdS5})-(\ref{rangeKerr5}), we have

\begin{eqnarray}
E^{\prime } &=&q(\partial _{t})=\frac{3\pi }{4G}\frac{m}{\Xi
_{a}\Xi _{b}},
\label{EprimeKerr5} \\
E_{0} &=&q_{0}(\partial _{t})=\frac{\pi \ell ^{2}}{96\Xi _{a}\Xi _{b}G}%
\left( 7\Xi _{a}\Xi _{b}+\Xi _{a}^{2}+\Xi _{b}^{2}\right) ,
\label{E0Kerr5AJ}
\end{eqnarray}%
and the angular momenta

\begin{eqnarray}
J_{a} &=&q(\partial _{\phi })=\frac{\pi }{2G}\frac{ma}{\Xi
_{a}^{2}\Xi _{b}},
\label{Jphi5} \\
J_{b} &=&q(\partial _{\psi })=\frac{\pi }{2G}\frac{mb}{\Xi _{a}\Xi
_{b}^{2}}. \label{Jpsi5}
\end{eqnarray}

The quantity (\ref{EprimeKerr5}) is the energy obtained by Awad
and Johnson in \cite{Aw-Jo} that, however, does not satisfy the
first law of black hole thermodynamics, as has been pointed out in
\cite{GibbonsKerr}. The expression for the vacuum energy $E_{0}$
is in agreement with \cite{Aw-Jo}, and it is equivalent to the
Papadimitriou-Skenderis result \cite{Pa-Sk1}

\begin{equation}
E_{0}=\frac{3\pi \ell ^{2}}{32G}\left( 1+\frac{\left( \Xi _{a}-\Xi
_{b}\right) ^{2}}{9\Xi _{a}\Xi _{b}}\right) ,  \label{E0Kerr5PS}
\end{equation}%
both computed using different versions of the counterterms method.

Notice that the physical energy for Kerr-AdS is obtained with a
Killing vector that does not rotate at infinity $\xi =\partial
_{t}+\frac{a}{\ell^{2}}\partial _{\phi }+\frac{b}{\ell ^{2}}\partial _{\psi }$,%

\begin{equation}
E=q(\partial _{t}+\frac{a}{\ell ^{2}}\partial _{\phi }+\frac{b}{\ell ^{2}}%
\partial _{\psi })=\frac{\pi m}{4G\Xi _{a}^{2}\Xi _{b}^{2}}\left( 2\Xi
_{a}+2\Xi _{b}-\Xi _{a}\Xi _{b}\right) .  \label{EKerr5}
\end{equation}

The on-shell Euclidean action is

\begin{equation}
I_{5}^{E}=\frac{\beta }{2\pi G\ell ^{2}}\int_{r_{+}}^{\infty
}dr\int d\Omega N\sqrt{-h}+\beta c_{4}\int d\Omega \left. \left(
B_{4}\right) \right\vert ^{r=\infty }
\end{equation}%
and we see that the boundary term cancels out the divergences
coming from the bulk action, such that we have the finite result

\begin{eqnarray}
I_{5}^{E} &=&\frac{\beta \pi }{96\ell ^{2}\Xi _{a}\Xi _{b}G}\left(
-a^{4}+9a^{2}\ell ^{2}+24r_{+}^{2}a^{2}+17a^{2}b^{2}-24m\ell
^{2}-9\ell
^{4}+24r_{+}^{4}\right. -  \notag \\
&&-b^{4}+9\ell ^{2}b^{2}+24r_{+}^{2}b^{2}),
\end{eqnarray}%

This expression can also be put into the form of Awad-Johnson

\begin{eqnarray}
I_{5}^{E} &=&\frac{\beta \pi }{96\ell ^{2}\Xi _{a}\Xi _{b}G}\left(
12\left( r_{+}^{2}/\ell ^{2}\right) (1-\Xi _{a}-\Xi _{b})+\Xi
_{a}^{2}+\Xi
_{b}^{2}+\Xi _{a}\Xi _{b}+12r_{+}^{4}/\ell ^{4}-\right.  \notag \\
&&\left. -2(a^{4}+b^{4})/\ell ^{4}-12\left( a^{2}b^{2}/\ell
^{4}\right) (\ell ^{2}/r_{+}^{2}-1/3)-12\right) ,
\end{eqnarray}%
or the more compact one obtained by Papadimitriou-Skenderis%

\begin{equation}
I_{5}^{E}=\beta E_{0}+\frac{\beta \pi }{4G\ell ^{2}\Xi _{a}\Xi
_{b}}(m\ell ^{2}-(r_{+}^{2}+a^{2})(r_{+}^{2}+b^{2})).
\end{equation}%

In order to obtain the correct value for the entropy, one must use $\mathcal{%
E}=E+E_{0}$ as the total energy for the black hole system, the angular
velocities respect to a non-rotating frame at infinity%

\begin{equation}
\Omega _{a}=\frac{a(1+r_{+}^{2}\ell
^{-2})}{r_{+}^{2}+a^{2}},\qquad \Omega
_{b}=\frac{b(1+r_{+}^{2}\ell ^{-2})}{r_{+}^{2}+b^{2}},
\label{omegaab}
\end{equation}%
and the Euclidean period%

\begin{equation}
\beta =\frac{2\pi (r_{+}^{2}+a^{2})(r_{+}^{2}+b^{2})\ell ^{2}}{%
2r_{+}^{6}+r_{+}^{4}(\ell ^{2}+b^{2}+a^{2})-a^{2}b^{2}\ell ^{2}}.
\label{betaKerr5}
\end{equation}%

The above quantities satisfy the thermodynamical relation%

\begin{equation}
S=\beta (\mathcal{E}-\Omega _{a}J_{a}-\Omega _{b}J_{b})-I_{5}^{E}=\frac{Area%
}{4G}
\end{equation}%
that recovers the entropy in terms of the area for Kerr-AdS black hole (\ref%
{areaKerr5}).

\textit{-Clarkson-Mann Solitons}

In recent papers \cite{Cla-Man1,Cla-Man2}, new solitons in
cosmological spacetimes were presented. These solutions resemble
the Eguchi-Hanson metrics in four dimensions \cite{Egu-Han} and in
the case of a negative cosmological constant, they posses $AdS/Zp$
asymptotics and a lower energy than the global AdS or even global
$AdS/Zp$ spacetimes.

The Clarkson-Mann-AdS soliton metric reads%

\begin{eqnarray}
ds^{2} &=&-g(r)dt^{2}+\frac{r^{2}f(r)}{4}[d\psi +\cos \theta d\phi ]^{2}+%
\frac{dr^{2}}{f(r)g(r)}+\frac{r^{2}}{4}d\Omega _{2}^{2},  \notag \\
g(r) &=&1+\frac{r^{2}}{\ell ^{2}},\qquad
f(r)=1-\frac{a^{4}}{r^{4}},
\end{eqnarray}%
where $d\Omega _{2}^{2}$ is the metric of the unit $2-$sphere.

In order to remove the stringlike singularity at $r=a$, the period of $\psi $
must be $4\pi /p$ and the parameter $a$ satisfies the relation%

\begin{equation}
a^{2}=\ell ^{2}\left( \frac{p^{2}}{4}-1\right) ,
\end{equation}%
with $p\geq 3$.

The energy for this solitonic solution is negative%

\begin{equation}
E=q(\partial _{t})=-\frac{\pi a^{4}}{8G\ell ^{2}p}
\label{EClaMan}
\end{equation}%
in agreement to the result computed using the standard counterterms
procedure in \cite{Cla-Man1,Ast-Man-Ste}. The present method also
reproduces the value
of the vacuum energy, which is lower than that of global AdS spacetime%

\begin{equation}
E_{0}=q_{0}(\partial _{t})=\frac{3\pi \ell ^{2}}{32Gp}.
\end{equation}%

The negative mass (\ref{EClaMan}) has also been found in
\cite{Ce-Sa-Te} through a spin-connection formulation of the
Abbott-Deser \cite{A-D} (and, more recently, Deser-Tekin
\cite{Des-Tek1,Des-Tek2}) method.

The total Euclidean action

\begin{equation}
I^{E}=I_{bulk}^{E}+c_{4}\int_{\partial {M}}B_{4}^{E},
\end{equation}%
turns out to be

\begin{equation}
I^{E}=\beta (E+E_{0})
\end{equation}%
where the Euclidean period $\beta $ remains arbitrary, as the
solution is horizonless. As a consequence, the entropy of the
system is zero.

\section{Seven-Dimensional Case}

Before we go into the general odd-dimensional case, let us consider $D=7$
also for illustrative purposes. The expression for the boundary term%

\begin{eqnarray}
B_{6} &=&-6\int_{0}^{1}dt\int_{0}^{t}ds\varepsilon
_{a_{1}...a_{6}}K^{a_{1}}e^{a_{2}}(R^{a_{3}a_{4}}-t^{2}K^{a_{3}}K^{a_{4}}+%
\frac{s^{2}}{\ell ^{2}}e^{a_{3}}e^{a_{4}})\times  \notag \\
&&\times (R^{a_{5}a_{6}}-t^{2}K^{a_{5}}K^{a_{6}}+\frac{s^{2}}{\ell ^{2}}%
e^{a_{5}}e^{a_{6}}),
\end{eqnarray}%
after the parametric integrations are performed is given by%

\begin{eqnarray}
B_{6} &=&-3\varepsilon
_{a_{1}...a_{6}}K^{a_{1}}e^{a_{2}}(R^{a_{3}a_{4}}R^{a_{5}a_{6}}-\frac{1}{3}%
K^{a_{3}}K^{a_{4}}K^{a_{5}}K^{a_{6}}+\frac{1}{15\ell ^{4}}%
e^{a_{3}}e^{a_{4}}e^{a_{5}}e^{a_{6}}  \notag \\
&&-R^{a_{3}a_{4}}K^{a_{5}}K^{a_{6}}+\frac{1}{3\ell ^{2}}%
R^{a_{3}a_{4}}e^{a_{5}}e^{a_{6}}-\frac{2}{9\ell ^{2}}%
K^{a_{3}}K^{a_{4}}e^{a_{5}}e^{a_{6}}).
\end{eqnarray}

The equivalent tensorial form of the Kounterterms for seven dimensions is

\begin{eqnarray}
B_{6} &=&\frac{3}{4}\sqrt{-h}\delta _{\lbrack
j_{1}...j_{5}]}^{[i_{1}...i_{5}]}K_{i_{1}}^{j_{1}}(R_{i_{2}i_{3}}^{j_{2}j_{3}}R_{i_{4}i_{5}}^{j_{4}j_{5}}-%
\frac{4}{3}%
K_{i_{2}}^{j_{2}}K_{i_{3}}^{j_{3}}K_{i_{4}}^{j_{4}}K_{i_{5}}^{j_{5}}+\frac{4%
}{15\ell ^{4}}\delta _{i_{2}}^{j_{2}}\delta _{i_{3}}^{j_{3}}\delta
_{i_{4}}^{j_{4}}\delta _{i_{5}}^{j_{5}}  \notag \\
&&-2R_{i_{2}i_{3}}^{j_{2}j_{3}}K_{i_{4}}^{j_{4}}K_{i_{5}}^{j_{5}}+\frac{2}{%
3\ell ^{2}}R_{i_{2}i_{3}}^{j_{2}j_{3}}\delta
_{i_{4}}^{j_{4}}\delta
_{i_{5}}^{j_{5}}-\frac{8}{9\ell ^{2}}K_{i_{2}}^{j_{2}}K_{i_{3}}^{j_{3}}%
\delta _{i_{4}}^{j_{4}}\delta _{i_{5}}^{j_{5}}).
\end{eqnarray}

\subsection{Action Principle}

Now we develop a similar treatment as in the five-dimensional
case, writing down an adequate form of the variation of the
boundary term, in order to use suitable boundary conditions for
AAdS spacetimes. Varying the
seven-dimensional action, we have --on-shell-- a total surface term%

\begin{eqnarray}
\delta I_{7} &=&-2\int_{\partial M}\kappa _{_{7}}\epsilon
_{abcdfg}\delta
K^{a}e^{b}e^{c}e^{d}e^{f}e^{g}+\frac{3}{2}c_{6}\epsilon
_{abcdfg}\delta
K^{a}e^{b}\Big(R^{cd}R^{fg}+\frac{5}{3}K^{c}K^{d}K^{f}K^{g}  \notag \\
&&+\frac{1}{15\ell ^{4}}e^{c}e^{d}e^{f}e^{g}-3R^{cd}K^{f}K^{g}+\frac{1}{%
3\ell ^{2}}R^{cd}e^{f}e^{g}-\frac{2}{3\ell ^{2}}K^{c}K^{d}e^{f}e^{g}\Big)+%
\frac{3}{2}c_{6}\epsilon _{abcdfg}K^{a}\delta
e^{b}\Big(R^{cd}R^{fg}  \notag
\\
&&+\frac{1}{3}K^{c}K^{d}K^{f}K^{g}+\frac{1}{3\ell ^{4}}%
e^{c}e^{d}e^{f}e^{g}-R^{cd}K^{f}K^{g}+\frac{1}{\ell ^{2}}R^{cd}e^{f}e^{g}-%
\frac{2}{3\ell ^{2}}K^{c}K^{d}e^{f}e^{g}\Big).
\end{eqnarray}%

The constant in front of the bulk action is $\kappa _{7}=1/(16\pi G\times 5!)
$. The total variation can be put into the form%

\begin{eqnarray}
\delta I_{7} &=&-2\int_{\partial M}\kappa _{_{7}}\epsilon
_{abcdfg}\delta
K^{a}e^{b}e^{c}e^{d}e^{f}e^{g}+3c_{6}\epsilon _{abcdfg}\delta K^{a}e^{b}\Big(%
\hat{R}^{cd}\hat{R}^{fg}+\frac{2}{3\ell ^{2}}\hat{R}^{cd}e^{f}e^{g}+\frac{1}{%
5\ell ^{4}}e^{c}e^{d}e^{f}e^{g}\Big)  \notag \\
&&-\frac{3}{2}c_{6}\epsilon _{abcdfg}(\delta K^{a}e^{b}-K^{a}\delta e^{b})%
\Big(R^{cd}R^{fg}+\frac{1}{3}K^{c}K^{d}K^{f}K^{g}+\frac{1}{3\ell ^{4}}%
e^{c}e^{d}e^{f}e^{g}-R^{cd}K^{f}K^{g}  \notag \\
&&+\frac{1}{\ell ^{2}}R^{cd}e^{f}e^{g}-\frac{2}{3\ell ^{2}}%
K^{c}K^{d}e^{f}e^{g}\Big).
\end{eqnarray}%
using again the Gauss-Coddazzi relation (\ref{GaCod}). The above
relation already hints a pattern for the surface term coming from
the total variation of the action,%

\begin{eqnarray}
\delta I_{7} &=&-2\int_{\partial M}\kappa _{_{7}}\epsilon
_{abcdfg}\delta
K^{a}e^{b}e^{c}e^{d}e^{f}e^{g}+3c_{6}\int_{0}^{1}dt\epsilon
_{abcdfg}\delta
K^{a}e^{b}\Big(\hat{R}^{cd}+\frac{t^{2}}{\ell ^{2}}e^{c}e^{d}\Big)\Big(\hat{R%
}^{fg}+\frac{t^{2}}{\ell ^{2}}e^{f}e^{g}\Big)  \notag \\
&&\!\!\!\!\!\!\!-3c_{6}\int_{0}^{1}dt\,t\epsilon _{abcdfg}(\delta
K^{a}e^{b}\!-\!K^{a}\delta e^{b})\Big(R^{cd}\!-\!t^{2}K^{c}K^{d}\!+\!\frac{%
t^{2}}{\ell ^{2}}e^{c}e^{d}\Big)\Big(R^{fg}\!-\!t^{2}K^{f}K^{g}\!+\!\frac{%
t^{2}}{\ell ^{2}}e^{f}e^{g}\Big)  \label{var7}
\end{eqnarray}%
that we will confirm below in the general odd-dimensional case.

As in the five-dimensional case, we consider a more explicit form
of the second line

\begin{eqnarray}
&&\varepsilon _{abcdfg}\,\varepsilon
^{i_{1}i_{2}i_{3}i_{4}i_{5}i_{6}}\left[ \delta
K_{i_{1}}^{j}e_{j}^{a}e_{i_{2}}^{b}+\delta
e_{j}^{a}e_{l}^{b}\left(
K_{i_{1}}^{j}\delta _{i_{2}}^{l}-K_{i_{2}}^{l}\delta _{i_{1}}^{j}\right) %
\right] \times  \notag \\
&&\times \int_{0}^{1}dt\,t\left( \frac{1}{2}%
R_{i_{3}i_{4}}^{cd}-t^{2}K_{i_{3}}^{c}K_{i_{4}}^{d}+\frac{t^{2}}{\ell ^{2}}%
\,e_{i_{3}}^{c}e_{i_{4}}^{d}\right) \left( \frac{1}{2}%
R_{i_{5}i_{6}}^{fg}-t^{2}K_{i_{5}}^{f}K_{i_{6}}^{g}+\frac{t^{2}}{\ell ^{2}}%
e_{i_{5}}^{f}e_{i_{6}}^{g}\right) d^{6}x,
\end{eqnarray}%
that again vanishes identically for the boundary condition on the
extrinsic curvature (\ref{Kdelta}) and the corresponding variation
(\ref{delKdelta}). Thus, the problem of a well-defined action
principle amounts to fixing the coupling of the boundary term
$c_{6}$. In the asymptotic region the spacetime curvature is
constant and then, inserting (\ref{ALAdSdf}) in the
first line of eq.(\ref{var7}), we have%

\begin{equation}
\delta I_{7}=-2\int_{\partial M}\epsilon _{abcdfg}\delta
K^{a}e^{b}e^{c}e^{d}e^{f}e^{g}\left( \kappa _{_{7}}+3\frac{c_{6}}{\ell ^{4}}%
\int_{0}^{1}dt(-1+t^{2})^{2}\right) .
\end{equation}%

Thus, the cancelation of the surface term implies%

\begin{equation}
c_{6}=-\frac{5}{8}\kappa _{_{7}}\ell ^{4}=-\frac{\ell ^{4}}{16\pi G\times 192%
}.
\end{equation}%

Now that we have achieved a well-posed action principle, we
discuss the construction of the Noether charges that derive from
it.

\subsection{Conserved Charges}

The Noether current for the seven-dimensional case is

\begin{equation}
\ast J=-\Theta (e^{a},K^{a},\delta e^{a},\delta K^{a})-i_{\xi
}(L_{7}+c_{6}dB_{6}),
\end{equation}%
where $L_{7}$ is the bulk Lagrangian and $\Theta $ is the surface term%

\begin{eqnarray}
\Theta  &=&-2\kappa _{7}\int_{\partial M}\varepsilon
_{abcdfg}\delta
K^{a}e^{b}\left[ e^{c}e^{d}e^{f}e^{g}-\frac{15}{8}\ell ^{4}\int_{0}^{1}dt%
\Big(\hat{R}^{cd}+\frac{t^{2}}{\ell ^{2}}e^{c}e^{d}\Big)\Big(\hat{R}^{fg}+%
\frac{t^{2}}{\ell ^{2}}e^{f}e^{g}\Big)\right]   \notag \\
&&\!\!\!\!\!\!\!\!\!\!\!\!\!\!\!\!\!\!+\frac{15}{8}\ell
^{4}\!\int_{0}^{1}dt\,t\varepsilon _{abcdfg}(\delta
K^{a}e^{b}\!-\!K^{a}\delta e^{b})\!\Big(R^{cd}\!-\!t^{2}K^{c}K^{d}\!+\!\frac{%
t^{2}}{\ell ^{2}}e^{c}e^{d}\Big)\!\Big(R^{fg}\!-\!t^{2}K^{f}K^{g}\!+\!\frac{%
t^{2}}{\ell ^{2}}e^{f}e^{g}\Big)  \label{theta7}
\end{eqnarray}%

Carrying out the construction in Appendix \ref{Noether} for a bulk
Lagrangian supplemented in a boundary term, the Noether charge is written as%

\begin{equation}
Q(\xi )=\int_{\partial \Sigma }2\kappa _{7}\varepsilon
_{abcdfg}\,i_{\xi
}K^{a}e^{b}e^{c}e^{d}e^{f}e^{g}+c_{6}\left( i_{\xi }K^{a}\frac{\delta B_{6}}{%
\delta K^{a}}+i_{\xi }e^{a}\frac{\delta B_{6}}{\delta
e^{a}}\right) \label{charge7}
\end{equation}%

The formula for the conserved quantities contains two contributions

\begin{equation}
Q(\xi )=q(\xi )+q_{0}(\xi ),
\end{equation}%
that can be traced back to the first and second lines in the surface term (%
\ref{theta7}). The first part

\begin{equation}
q(\xi )=2\kappa _{7}\int_{\partial \Sigma }\varepsilon
_{abcdfg}i_{\xi
}K^{a}e^{b}\left[ e^{c}e^{d}e^{f}e^{g}-\frac{15}{8}\ell ^{4}\int_{0}^{1}dt%
\Big(\hat{R}^{cd}+\frac{t^{2}}{\ell ^{2}}e^{c}e^{d}\Big)\Big(\hat{R}^{fg}+%
\frac{t^{2}}{\ell ^{2}}e^{f}e^{g}\Big)\right] ,  \label{qxi7}
\end{equation}%
will provide the mass and angular momentum for AAdS solutions.
Equivalently, eq.(\ref{qxi7}) can be factorized as

\begin{equation}
q(\xi )=-\frac{30}{8}\kappa _{7}\ell ^{4}\int_{\partial \Sigma
}\varepsilon
_{abcdfg}i_{\xi }K^{a}e^{b}\Big(\hat{R}^{cd}+\frac{1}{\ell ^{2}}e^{c}e^{d}%
\Big)\Big(\hat{R}^{fg}-\frac{1}{3\ell ^{2}}e^{f}e^{g}\Big)
\end{equation}%
that in its tensorial form%

\begin{equation}
q(\xi )=-\frac{\ell ^{4}}{16\pi G\times 128}\int_{\partial \Sigma }\sqrt{-h}%
\varepsilon _{i_{1}...i_{6}}(\xi ^{k}K_{k}^{i_{1}})\delta _{j_{2}}^{i_{2}}%
\Big(\hat{R}_{j_{3}j_{4}}^{i_{3}i_{4}}+\frac{1}{\ell ^{2}}\delta
_{\lbrack
j_{3}j_{4}]}^{[i_{3}i_{4}]}\Big)\Big(\hat{R}_{j_{5}j_{6}}^{i_{5}i_{6}}-\frac{%
1}{3\ell ^{2}}\delta _{\lbrack j_{5}j_{6}]}^{[i_{5}i_{6}]}\Big)%
dx^{j_{2}}...dx^{j_{6}},  \label{qxi7tensor}
\end{equation}%
is proportional to the curvature for the AdS group and, therefore,
identically vanishing for (global) AdS spacetime. The product $%
dx^{j_{2}}\wedge ...\wedge dx^{j_{6}}$ is the volume element of
the boundary of the spatial section $\partial \Sigma $ (at
constant time).

As a consequence, the additional term $q_{0}$ in the conserved
quantities is responsible for the existence of a vacuum energy

\begin{eqnarray}
q_{0}(\xi ) &=&\frac{30}{8}\kappa _{7}\ell ^{4}\int_{\partial
\Sigma }\int_{0}^{1}dt\,t\varepsilon _{abcdfg}\left( i_{\xi
}K^{a}e^{b}+K^{a}i_{\xi
}e^{b}\right) \Big(R^{cd}-t^{2}K^{c}K^{d}+\frac{t^{2}}{\ell ^{2}}e^{c}e^{d}%
\Big)\times  \notag \\
&&\times \Big(R^{fg}-t^{2}K^{f}K^{g}+\frac{t^{2}}{\ell
^{2}}e^{f}e^{g}\Big),
\notag \\
&=&\frac{\ell ^{4}}{16\pi G\times 128}\int_{\partial \Sigma
}\int_{0}^{1}dt\,t\sqrt{-h}\varepsilon _{i_{1}...i_{6}}\xi
^{k}(K_{k}^{i_{1}}\delta _{j_{2}}^{i_{2}}+K_{j_{2}}^{i_{1}}\delta
_{k}^{i_{2}})\Big(%
R_{j_{3}j_{4}}^{i_{3}i_{4}}-t^{2}K_{[j_{3}j_{4}]}^{[i_{3}i_{4}]}+\frac{t^{2}%
}{\ell ^{2}}\delta _{\lbrack
j_{3}j_{4}]}^{[i_{3}i_{4}]}\Big)\times  \notag
\\
&&\times \Big(%
R_{j_{5}j_{6}}^{i_{5}i_{6}}-t^{2}K_{[j_{5}j_{6}]}^{[i_{5}i_{6}]}+\frac{t^{2}%
}{\ell ^{2}}\delta _{\lbrack j_{5}j_{6}]}^{[i_{5}i_{6}]}\Big)%
dx^{j_{2}}...dx^{j_{6}}  \label{q0xitensor7}
\end{eqnarray}%
where we have kept the parametric integral because, otherwise, the
expression is more involved.

\subsection{Examples}

-\textit{Topological Static Black Holes}

For the seven-dimensional static metric specified by eqs.(\ref{Sch-AdS})-(%
\ref{Delta2}), we compute the mass and the vacuum energy using expressions (%
\ref{qxi7tensor}) and (\ref{q0xitensor7}), respectively, for the
Killing vector $\xi =\partial _{t}$. Using the relations
(\ref{KSch})-(\ref{sqrth}), we obtain

\begin{eqnarray}
q(\partial _{t}) &=&M=5!Vol(\Sigma _{5}^{k})\lim_{r\rightarrow
\infty
}(\Delta ^{2})^{\prime }r\Big\{\kappa _{_{7}}[r^{4}+3c_{6}\int_{0}^{1}dt\Big(%
k-\Delta ^{2}+t^{2}\frac{r^{2}}{\ell ^{2}}\Big)^{2}\Big\}\!
\label{Mass7inter} \\
&=&\frac{Vol(\Sigma _{5}^{k})}{Vol(S^{5})}\mu .  \label{Mass7}
\end{eqnarray}%
for the mass, whereas for the vacuum energy takes the negative
value

\begin{eqnarray}
q_{0}(\partial _{t}) &=&E_{0}=-2\times 5!c_{6}Vol(\Sigma
_{5}^{k})\lim_{r\rightarrow \infty }\Big(\Delta
^{2}-\frac{r(\Delta
^{2})^{\prime }}{2}\Big)\int_{0}^{1}dt\,t\Big(k-t^{2}\Delta ^{2}+t^{2}\frac{%
r^{2}}{\ell ^{2}}\Big)^{2}  \label{E07inter} \\
&=&(-k)^{3}\frac{5\ell ^{4}}{128\pi G}Vol(\Sigma _{5}^{k}).
\label{E07}
\end{eqnarray}%

We have included an intermediate step in the computation of both
the mass and vacuum energy, because these expressions will appear
again in the evaluation of the Euclidean action.

With the Euclidean period $\beta $ defined as in
eq.(\ref{betaSAdS}), the
Euclidean bulk action, evaluated for a static black hole of the form (\ref%
{Sch-AdS}) in seven dimensions is

\begin{equation}
I_{bulk}^{E}=-5!\kappa _{_{7}}Vol(\Sigma _{5}^{k})\beta \left.
\lbrack (\Delta ^{2})^{\prime }r^{5}]\right\vert _{r_{+}}^{\infty
}.  \label{IEbulk7}
\end{equation}%

The boundary is defined only at the asymptotic region and then,
the Euclidean boundary term is

\begin{eqnarray}
\int_{\partial {M}}B_{6}^{E} &=&2\times 5!Vol(\Sigma _{5}^{k})\beta \Big[%
\frac{r(\Delta ^{2})^{\prime }}{2}\int_{0}^{1}dt\Big(k-\Delta ^{2}+\frac{%
t^{2}r^{2}}{\ell ^{2}}\Big)^{2}+  \notag \\
&&+\Big(\Delta ^{2}-\frac{r(\Delta ^{2})^{\prime }}{2}\Big)\int_{0}^{1}dt\,t%
\Big(k-t^{2}\Delta ^{2}+t^{2}\frac{r^{2}}{\ell ^{2}}\Big)^{2}\Big]\Big|%
^{\infty }.  \label{BE6}
\end{eqnarray}%
such that in the total Euclidean action

\begin{equation}
I_{7}^{E}=I_{bulk}^{E}+c_{6}\int_{\partial M}B_{6}^{E}
\end{equation}%
the contribution at $r=\infty $ can be read off from eqs.(\ref{Mass7inter}),(%
\ref{E07inter}) as $-\beta (M+E_{0})$ and%

\begin{equation}
I_{7}^{E}=\frac{Vol(\Sigma _{5}^{k})}{16\pi G}\beta
r_{+}^{5}\left. (\Delta ^{2})^{\prime }\right\vert _{r_{+}}-\beta
(M+E_{0}).
\end{equation}

Using the definition of thermodynamic energy $\mathcal{E}$ in eq.(\ref%
{thermoenergy}) --that is equivalent to the result obtained from
the Noether theorem-- we obtain the entropy as the Euclidean bulk action evaluated at $%
r=r_{+}$%

\begin{equation}
S=\frac{Vol(\Sigma _{5}^{k})r_{+}^{5}}{4G}=\frac{Area}{4G}.
\end{equation}

-\textit{Kerr-AdS}$_{7}$\textit{\ Black Hole}

The number of independent rotation parameters for $D-$dimensional Kerr-AdS
metric is equal to the number of Casimir invariants for the rotation group $%
SO(D-1)$, which is the integer part of $(D-1)/2$. The general
rotating black hole in seven-dimensional AdS gravity then
possesses three rotation parameters, but we shall consider below
the particular case of a single rotation parameter to show the
finiteness of the conserved quantities and Euclidean action.

The line element for the one-parameter Kerr-AdS$_{7}$ spacetime is

\begin{eqnarray}
ds^{2} &=&-\frac{\Delta _{r}}{\rho ^{2}}\left( dt-\frac{a\sin ^{2}\theta }{%
\Xi }d\phi \right) ^{2}+r^{2}\cos ^{2}\theta d\psi ^{2}+\frac{\rho ^{2}dr^{2}%
}{\Delta _{r}}+\frac{\rho ^{2}d\theta ^{2}}{\Delta _{\theta }}+  \notag \\
&&+\frac{\Delta _{\theta }\sin ^{2}\theta }{\rho ^{2}}\left( adt-\frac{%
\left( r^{2}+a^{2}\right) }{\Xi }d\phi \right) ^{2}+r^{2}\cos
^{2}\theta d\Omega _{3}^{2},  \label{KerrAdS7single}
\end{eqnarray}%
where the functions in the metric are%

\begin{eqnarray}
\Delta _{r} &=&\left( r^{2}+a^{2}\right) \left( 1+r^{2}/\ell
^{2}\right)
-2m/r^{2},  \label{Deltar7} \\
\Delta _{\theta } &\equiv &1-\frac{a^{2}}{\ell ^{2}}\cos
^{2}\theta ,
\label{Deltatheta} \\
\rho ^{2} &\equiv &r^{2}+a^{2}\cos ^{2}\theta ,\qquad \Xi \equiv 1-\frac{%
a^{2}}{\ell ^{2}},  \label{rho}
\end{eqnarray}%
and $d\Omega _{3}^{2}$ is the metric of the $3-$sphere%

\begin{equation}
d\Omega _{3}^{2}=d\psi ^{2}+\sin ^{2}\psi d\eta ^{2}+\cos ^{2}\psi
d\beta ^{2}  \label{3sph}
\end{equation}%
and the angles range is $\theta ,\psi \in \lbrack 0,\pi /2]$ and
$\phi ,\eta ,\beta \in \lbrack 0,2\pi ]$.

The area of the event horizon is

\begin{equation}
Area=\pi ^{3}\frac{r_{+}^{3}(r_{+}^{2}+a^{2})}{\Xi }.
\label{areaKerr7single}
\end{equation}%

Using the charge formulas (\ref{qxi7tensor}) and
(\ref{q0xitensor7}) for the metric
(\ref{KerrAdS7single})-(\ref{3sph}) yields

\begin{eqnarray}
E^{\prime } &=&q(\partial _{t})=\frac{5\pi ^{2}}{8G}\frac{m}{\Xi
},
\label{EprimeKerr7} \\
E_{0} &=&q_{0}(\partial _{t})=-\frac{\pi ^{2}}{1280\Xi G\ell
^{2}}\left( 50\ell ^{6}-50a^{2}\ell ^{4}+5a^{4}\ell
^{2}+a^{6}\right) ,  \label{E0Kerr7}
\end{eqnarray}%
and the angular momentum

\begin{equation}
J=q(\partial _{\phi })=\frac{\pi ^{2}}{4G}\frac{ma}{\Xi ^{2}}.
\label{Jpsi7}
\end{equation}%

The values in eqs. (\ref{EprimeKerr7}), (\ref{E0Kerr7}) and
(\ref{Jpsi7}) agree with the ones computed using a Dirichlet
counterterms regularization \cite{Das-Mann,Aw-Jo}. The vacuum
energy can be also written as

\begin{equation}
E_{0}=-\frac{5\pi ^{2}\ell ^{4}}{128G}\left( 1+\frac{\left( 1-\Xi
\right) ^{2}\left( 6-\Xi \right) }{50\Xi }\right)
\label{E07KerrOl}
\end{equation}%
in order to make more manifest the matching with the vacuum energy
for
Schwarzchild-AdS (\ref{E07}) in the non-rotating limit\footnote{%
The volume of $S^{5}$ is $\pi ^{3}$.} and to try to infer the
general vacuum energy in the case of three different rotation
parameters.

As pointed out in the five-dimensional section above, the physical
energy for a Kerr-AdS black hole is the conserved quantity
associated to a non-rotating asymptotic timelike Killing vector $\xi =\partial _{t}+\frac{a}{%
\ell ^{2}}\partial _{\phi }$ \cite{GibbonsKerr}%

\begin{equation}
E=q(\partial _{t}+\frac{a}{\ell ^{2}}\partial _{\phi })=\frac{m\pi ^{2}}{%
8G\Xi ^{2}}\left( 2+3\Xi \right) ,  \label{EKerr7one}
\end{equation}%
in agreement with the formulas in Refs.\cite%
{GibbonsKerr,Der-Kat,Barnich-Compere,Des-TekKerr}, specialized to
a single nonvanishing rotation parameter.

In order to complete the discussion about regularization of this
seven-dimensional solution, we compute the on-shell Euclidean
action, that for an stationary spacetime is given by

\begin{equation}
I_{7}^{E}=\frac{3\beta }{4\pi G\ell ^{2}}\int_{r_{+}}^{\infty
}dr\int d\Omega N\sqrt{h}+\beta c_{6}\int d\Omega \left. \left(
B_{6}\right) \right\vert ^{r=\infty },  \label{IE7st}
\end{equation}%
where the Euclidean period is%

\begin{equation}
\beta =\frac{2\pi (r_{+}^{2}+a^{2})r_{+}}{3r_{+}^{4}/\ell
^{2}+2r_{+}^{2}(1+a^{2}/\ell ^{2})+a^{2}}.  \label{betaKerr7}
\end{equation}%

The divergences at radial infinity in the bulk action are exactly
canceled by the ones in the Euclidean boundary term, such that we
get the finite result

\begin{equation}
I_{7}^{E}=-\frac{\beta \pi ^{2}\ell ^{4}}{1280\Xi G}\left[ 160\left( \frac{%
r_{+}^{4}a^{2}}{\ell ^{6}}+\frac{r_{+}^{6}}{\ell ^{6}}-\frac{m}{\ell ^{4}}%
\right) +\frac{a^{6}}{\ell ^{6}}+5\frac{a^{4}}{\ell ^{4}}+50\Xi
\right] ,
\end{equation}%
that can be conveniently rewritten as

\begin{equation}
I_{7}^{E}=\beta E_{0}+I_{7}^{\prime E},  \label{IE7Kount}
\end{equation}%
where%

\begin{equation}
I_{7}^{\prime E}=\frac{\beta \pi ^{2}}{8G\ell ^{2}\Xi }(m\ell
^{2}-r_{+}^{4}(r_{+}^{2}+a^{2}))  \label{IE7Kerrback}
\end{equation}%
corresponds to the value of the Euclidean action computed in a
background-substraction method \cite{GibbonsKerr}.
Eq.(\ref{IE7Kerrback}) satisfies the thermodynamical relation

\begin{equation}
S=\beta (E-\Omega J)-I_{7}^{\prime E}=\frac{Area}{4G},
\end{equation}%
for the energy (\ref{EKerr7one}), angular momentum (\ref{Jpsi7}),
the
angular velocity respect to a non-rotating frame at infinity%

\begin{equation}
\Omega =\frac{a(1+r_{+}^{2}\ell ^{-2})}{r_{+}^{2}+a^{2}},
\label{omegaa}
\end{equation}%
and the area of the event horizon%

\begin{equation}
Area=\pi ^{3}\frac{r_{+}^{3}(r_{+}^{2}+a^{2})}{\Xi }.
\label{AreaKerr7}
\end{equation}%

In turn, the Euclidean action (\ref{IE7Kount}), computed using the
background-independent Kounterterms prescription, obeys%

\begin{equation}
S=\beta (\mathcal{E}-\Omega J)-I_{7}^{E}=\frac{Area}{4G},
\end{equation}%
for a thermodynamical energy consistently shifted in the vacuum energy $%
\mathcal{E}=E+E_{0}$.

\section{General Odd-Dimensional Case}

We have already introduced the general form the Kounterterms series
adopts in any odd dimension $D=2n+1$,
eqs.(\ref{B2ntheta})-(\ref{B2ntensor}). The parametric integrations
provide the relative coefficients of the boundary terms when
eq.(\ref{B2ntensor}) is expanded as a polynomial in the extrinsic
and intrinsic curvature
\begin{equation}
B_{2n}=n!\sqrt{-h}\sum_{p=0}^{n-1}\frac{\left( 2n-2p-3\right)
!!}{\ell ^{2(n-1-p)}}b_{2n}^{(p)},  \label{B2n-b2n}
\end{equation}%
where%
\begin{equation}
b_{2n}^{(p)}=\,\delta _{\lbrack j_{1}\cdots j_{2p+1}]}^{[i_{1}\cdots
i_{2p+1}]}\sum_{q=0}^{p}\frac{(-1)^{p-q}\,}{(p-q)!q!}\frac{2^{n-(p+q+1)}}{%
(n-q)}\,R_{i_{1}i_{2}}^{j_{1}j_{2}}\cdots
R_{i_{2q-1}i_{2q}}^{j_{2q-1}j_{2q}}\,K_{i_{2q+1}}^{j_{2q+1}}\cdots
K_{i_{2p+1}}^{j_{2p+1}}\,.  \label{b2np}
\end{equation}

The surface term obtained from an arbitrary variation of the action (\ref%
{IgD})--or equivalently, (\ref{EHD})-- has a more involved form in
the general odd-dimensional case. In the Appendix \ref{varB2n}, we
summarize the process of variation of the action in the general
case, cast in an appropriate form that allows us to impose the
asymptotic conditions for AAdS spacetimes discussed above

\begin{eqnarray}
\delta I_{2n+1} &=&-2\int_{\partial M}\epsilon
_{a_{1}...a_{2n}}\delta K^{a_{1}}e^{a_{2}}\left[ \kappa
_{_{D}}e^{a_{3}}...e^{a_{2n}}+nc_{2n}\int_{0}^{1}dt\Big(\hat{R}^{a_{3}a_{4}}+%
\frac{t^{2}}{\ell ^{2}}e^{a_{3}}e^{a_{4}}\Big)\times ...\right.  \notag \\
&&\qquad \qquad \qquad \qquad \qquad \left. \times \Big(\hat{R}%
^{a_{2n-1}a_{2n}}+\frac{t^{2}}{\ell
^{2}}e^{a_{2n-1}}e^{a_{2n}}\Big)\right]
\notag \\
&&-nc_{2n}\int_{0}^{1}dt\,t\epsilon _{a_{1}...a_{2n}}(\delta
K^{a_{1}}e^{a_{2}}-K^{a_{1}}\delta e^{a_{2}})\Big(%
R^{a_{3}a_{4}}-t^{2}K^{a_{3}}K^{a_{4}}+\frac{t^{2}}{\ell ^{2}}%
e^{a_{3}}e^{a_{4}}\Big)\times ...  \notag \\
&&\qquad \qquad \qquad \qquad \times \Big(%
R^{a_{2n-1}a_{2n}}-t^{2}K^{a_{2n-1}}K^{a_{2n}}+\frac{t^{2}}{\ell ^{2}}%
e^{a_{2n-1}}e^{a_{2n}}\Big).  \label{delI2n+1}
\end{eqnarray}%

As we have already seen in the five and seven-dimensional cases,
no matter the terms that multiply the curl $\epsilon
_{a_{1}a_{2}...a_{2n}}(\delta
K^{a_{1}}e^{a_{2}}-K^{a_{1}}\delta e^{a_{2}})$, the asymptotic conditions (%
\ref{Kdelta}) and (\ref{delKdelta}) cancels it identically. The problem of
making the full action stationary then reduces to fix the coupling constant $%
c_{2n}$ of the boundary term. This can be done demanding the
spacetime to be of constant curvature at the asymptotic region (eq.(\ref%
{ALAdSdf})). In a Riemannian manifold, this condition is
equivalent to asymptotic flatness for the curvature of the AdS
group

\begin{equation}
F=dA+A\wedge A=\frac{1}{2}\left( \hat{R}^{AB}+\frac{e^{A}e^{B}}{\ell ^{2}}%
\right) J_{AB}+\frac{T^{A}}{\ell }P_{A},
\end{equation}%
where $A=\frac{1}{2}\omega ^{AB}J_{AB}+\frac{e^{A}}{\ell }P_{A}$ is the $%
SO(D-1,2)$ group connection field, $\{J_{AB},P_{A}\}$ are the
generators of AdS rotations and translations, respectively, and
$T^{A}=\frac{1}{2}T_{\mu \nu }^{A}dx^{\mu }\wedge dx^{\nu }$ is
the two-form torsion.

Assuming (\ref{ALAdSdf}), we obtain the value

\begin{eqnarray}
c_{2n} &=&-\ell ^{2n-2}\frac{\kappa _{_{D}}}{n}\Big[%
\int_{0}^{1}dt(t^{2}-1)^{n-1}\Big]^{-1}, \\
&=&-\frac{(-\ell ^{2})^{n-1}}{4^{n+1}\pi G \,n\left[ (n-1)!%
\right] ^{2}}
\end{eqnarray}%
whose fixing is equivalent to canceling the highest-order
divergences in the Euclidean action.

\subsection{Noether Charges}

The conserved current associated to an isometry and prescribed by
the Noether theorem, in the general odd-dimensional case is

\begin{equation}
\ast J=-\Theta (e^{a},K^{a},\delta e^{a},\delta K^{a})-i_{\xi
}(L_{2n+1}+c_{2n}dB_{2n}).  \label{J2n+1}
\end{equation}%

Here, $L_{2n+1}$ is the bulk Lagrangian in $2n+1$ dimensions,
$B_{2n}$ is the regularizing Kounterterms series and $\Theta $ is
the surface term in the variation of the action (\ref{delI2n+1}).

Either replacing the explicit form of the terms in
eq.(\ref{J2n+1}) to write down the Noether current as $\ast
J=dQ(\xi )$ or employing the shortcut to
the charge derivation in Appendix \ref{Noether}%

\begin{equation}
Q(\xi )=\int_{\partial \Sigma }2\kappa _{D}\varepsilon
_{a_{1}...a_{2n}}\,i_{\xi
}K^{a_{1}}e^{a_{2}}...e^{a_{2n}}+c_{2n}\left( i_{\xi
}K^{a}\frac{\delta B_{2n}}{\delta K^{a}}+i_{\xi }e^{a}\frac{\delta
B_{2n}}{\delta e^{a}}\right)
\end{equation}%
the conserved charge is split as

\begin{equation}
Q(\xi )=q(\xi )+q_{0}(\xi ). \label{Q2n+1}
\end{equation}%

The first contribution can be traced back to the first two lines
in the surface term (\ref{delI2n+1})

\begin{eqnarray}
q(\xi ) &=&\frac{1}{2^{n-2}}\int_{\partial \Sigma }\kappa _{_{D}}\sqrt{-h}%
\epsilon _{i_{1}...i_{2n}}(\xi ^{k}K_{k}^{i_{1}})\delta _{j_{2}}^{i_{2}}%
\left[ \delta _{\lbrack j_{3}j_{4}]}^{[i_{3}i_{4}]}...\delta
_{\lbrack
j_{2n-1}j_{2n}]}^{[i_{2n-1}i_{2n}]}+\right.  \notag \\
&&\left. +nc_{2n}\int_{0}^{1}dt\Big(\hat{R}_{j_{3}j_{4}}^{i_{3}i_{4}}+\frac{%
t^{2}}{\ell ^{2}}\delta _{\lbrack j_{3}j_{4}]}^{[i_{3}i_{4}]}\Big)...\Big(%
\hat{R}_{j_{2n-1}j_{2n}}^{i_{2n-1}i_{2n}}+\frac{t^{2}}{\ell
^{2}}\delta _{\lbrack
j_{2n-1}j_{2n}]}^{[i_{2n-1}i_{2n}]}\Big)\right]
dx^{j_{2}}...dx^{j_{2n}},  \label{qtensor2n+1}
\end{eqnarray}%
and the second one comes from the curl term in the same term%

\begin{eqnarray}
q_{0}(\xi ) &=&-\frac{nc_{2n}}{2^{n-2}}\int_{\partial \Sigma }\sqrt{-h}%
\int_{0}^{1}dt\,t\epsilon _{i_{1}i_{2}...i_{2n}}\xi ^{k}(\delta
_{j_{2}}^{i_{2}}K_{k}^{i_{1}}+\delta _{k}^{i_{2}}K_{j_{2}}^{i_{1}})\Big(%
R_{j_{3}j_{4}}^{i_{3}i_{4}}-t^{2}K_{[j_{3}j_{4}]}^{[i_{3}i_{4}]}+\frac{t^{2}%
}{\ell ^{2}}\delta _{\lbrack j_{3}j_{4}]}^{[i_{3}i_{4}]}\Big)...  \notag \\
&&\Big(%
R_{j_{2n-1}j_{2n}}^{i_{2n-1}i_{2n}}-t^{2}K_{[j_{2n-1}j_{2n}]}^{[i_{2n-1}i_{2n}]}+%
\frac{t^{2}}{\ell ^{2}}\delta _{\lbrack j_{2n-1}j_{2n}]}^{[i_{2n-1}i_{2n}]}%
\Big)dx^{j_{2}}dx^{j_{3}}...dx^{j_{2n}}.  \label{q0tensor2n+1}
\end{eqnarray}

Evaluating the formula (\ref{qtensor2n+1}) for $\xi =\partial
_{t}$ in AAdS static black holes (\ref{Sch-AdS})-(\ref{Delta2})
gives

\begin{equation}
q(\partial _{t})=(D-2)!Vol(\Sigma _{D-2}^{k})\lim_{r\rightarrow
\infty
}(\Delta ^{2})^{\prime }r\Big\{\kappa _{_{D}}r^{2(n-1)}+nc_{2n}\int_{0}^{1}dt%
\Big(k-\Delta ^{2}+\frac{t^{2}r^{2}}{\ell ^{2}}\Big)^{n-1}\Big\},
\label{massgendel}
\end{equation}%
where the derivative of the function in the metric is%

\begin{equation}
(\Delta ^{2})^{\prime }r=2(D-3)\frac{\omega _{D}G\mu }{r^{D-3}}+2\frac{r^{2}%
}{\ell ^{2}}.
\end{equation}%

Expanding the second term as%

\begin{equation}
nc_{2n}\int_{0}^{1}dt\Big(k-\Delta ^{2}+\frac{t^{2}r^{2}}{\ell ^{2}}\Big)%
^{n-1}=\kappa _{D}\left( -r^{2(n-1)}+(2n-1)\omega _{D}G\frac{\ell ^{2}}{r^{2}%
}\mu +...\right)
\end{equation}%
where the additional contributions of lower order in $r$ are
irrelevant in
the limit $r\rightarrow \infty $, the topological black hole mass is finally%

\begin{equation*}
q(\partial _{t})=M=\frac{Vol(\Sigma _{D-2}^{k})}{Vol(S^{D-2})}\mu
.
\end{equation*}

The other formula in the conserved charge, $q_{0}(\xi )$,
specialized for a timelike Killing vector and topological black
holes, produces

\begin{equation}
q_{0}(\partial _{t})=-2c_{2n}(D-2)!Vol(\Sigma
_{D-2}^{k})\lim_{r\rightarrow
\infty }\Big(\Delta ^{2}-\frac{r(\Delta ^{2})^{\prime }}{2}\Big)%
\int_{0}^{1}dt\,t\Big(k-t^{2}\Delta ^{2}+\frac{t^{2}r^{2}}{\ell ^{2}}\Big)%
^{n-1},  \label{vac}
\end{equation}%
where the explicit evaluation of

\begin{equation}
\Big(\Delta ^{2}-\frac{r(\Delta ^{2})^{\prime
}}{2}\Big)=k-(D-1)\frac{\omega _{D}G}{r^{D-3}}\mu ,
\end{equation}%

\begin{equation}
\Big(k-t^{2}\Delta ^{2}+\frac{t^{2}r^{2}}{\ell ^{2}}\Big)=k(1-t^{2})+2t^{2}%
\frac{\omega _{D}G}{r^{D-3}}\mu ,
\end{equation}%
introduces at most finite contributions to the zero-point (vacuum) energy%

\begin{equation}
q_{0}(\partial _{t})=E_{0}=(-k)^{n}\frac{Vol(\Sigma
_{D-2}^{k})}{8\pi G}\ell ^{2n-2}\frac{(2n-1)!!^{2}}{(2n)!}.
\end{equation}%

This expression corroborates the general formula for the vacuum energy for
odd-dimensional AdS spacetime conjectured in Ref.\cite{Emparan-Johnson-Myers}%
, based on an extrapolation of explicit results in the
counterterms method up to nine dimensions.

In order to verify the consistency of the black hole thermodynamics, we
compute the total Euclidean action%

\begin{equation}
I_{2n+1}^{E}=I_{bulk}^{E}+c_{2n}\int_{\partial M}B_{2n}^{E},
\end{equation}%
for SAdS black hole. The bulk term is a total derivative, such
that the integration in the radial coordinate in the interval
$[r_{+},\infty )$ is simply

\begin{equation}
I_{bulk}^{E}=-\kappa _{D}(D-2)!Vol(\Sigma _{D-2}^{k})\beta
\{(\Delta ^{2})^{\prime }r^{D-2}\}|_{r_{+}}^{\infty },
\label{IEbulk2n+1}
\end{equation}%
and the Euclidean boundary term is

\begin{eqnarray}
\int_{\partial {M}}B_{2n}^{E} &=&2(D-2)!Vol(\Sigma _{D-2}^{k})\beta \Big[%
\frac{r(\Delta ^{2})^{\prime }}{2}\int_{0}^{1}dt\Big(k-\Delta ^{2}+\frac{%
t^{2}r^{2}}{\ell ^{2}}\Big)^{n-1}+ \\
&&+\Big(\Delta ^{2}-\frac{r(\Delta ^{2})^{\prime }}{2}\Big)\int_{0}^{1}dt\,t%
\Big(k-t^{2}\Delta ^{2}+\frac{t^{2}r^{2}}{\ell ^{2}}\Big)^{n-1}\Big]\Big|%
^{\infty }.
\end{eqnarray}%

The total contribution at $r=\infty $ can be identify as $-\beta (M+E_{0})$
and then the total Euclidean action is%

\begin{equation}
I_{2n+1}^{E}=\frac{Vol(\Sigma _{D-2}^{k})}{16\pi G}\beta
r_{+}^{D-2}\left. (\Delta ^{2})^{\prime }\right\vert
_{r_{+}}-\beta (M+E_{0}).
\end{equation}

The definition of thermodynamic energy $\mathcal{E}$ in eq.(\ref%
{thermoenergy}) recovers the total energy in the Noether theorem
$Q(\partial _{t})=q(\partial _{t})+q_{0}(\partial _{t})$, such
that eq.(\ref{SIE}) implies an entropy for SAdS black hole%

\begin{equation}
S=\frac{Vol(\Sigma _{D-2}^{k})r_{+}^{D-2}}{4G}=\frac{Area}{4G}.
\end{equation}

\section{Conclusions}

In this paper, we have explicitly shown the tensorial form of the
Kounterterms that regularize the action for AdS gravity in all odd
dimensions.

The key point of the construction is a well-principle action
principle that respects boundary conditions consistent with the
asymptotic behavior of a generic AAdS spacetime.

A definite form of the boundary terms achieves a finite action
principle: the action is stationary under arbitrary variations of
the fields and the conserved charges and the Euclidean action are
finite.

In the general odd-dimensional case, we have not computed the
conserved quantities in AAdS solutions other than for topological
SAdS black holes.

Further evaluation of more complex solutions will be certainly
more involved, but it could also reveal the general form that the
vacuum energy adopts for certain AAdS spaces. Particularly
interesting could be extending the results of the vacuum energy in
Refs.\cite{Aw-Jo,Pa-Sk2} to the recently generalized
multi-parameter Kerr-AdS black hole \cite{GibbonsKerr}. It is also
worthwhile to notice that the existing results on vacuum energy
for Kerr-AdS solutions, reproduced here, have been critically revised in Ref.%
\cite{GibbonsCasimir} because of their explicit dependence on the
rotation parameters. There, it is claimed that, in the same way
the energy and the angular velocities are referred to a coordinate
frame that is nonrotating at infinity, one should choose the
Einstein static universe as the metric on the conformal boundary
of the rotating black hole. As a consequence, the corresponding
Casimir energy is genuinely a constant and matches the one for
SAdS black hole.

But, how can we be sure that $q_{0}(\partial _{t})$ would always
produce the vacuum energy?

The reasoning is quite simple and has to do with the general form
of the other quantity that enters in the Noether charge, $q(\xi
)$. It can be proved that eq.(\ref{qtensor2n+1}) can be factorized by the l.h.s. of eq.(%
\ref{Reqdelta}), in the corresponding boundary indices%

\begin{equation}
q(\xi )=\frac{nc_{2n}}{2^{n-2}}\int_{\partial \Sigma
}\sqrt{-h}\epsilon
_{i_{1}...i_{2n}}(\xi ^{k}K_{k}^{i_{1}})\delta _{j_{2}}^{i_{2}}\Big(\hat{R}%
_{j_{3}j_{4}}^{i_{3}i_{4}}+\frac{1}{\ell ^{2}}\delta _{\lbrack
j_{3}j_{4}]}^{[i_{3}i_{4}]}\Big)\mathcal{P}%
_{j_{5}...j_{2n}}^{i_{5}...i_{2n}}dx^{j_{2}}...dx^{j_{2n}},
\end{equation}%
where $\mathcal{P}$ is a Lovelock-type polynomial of $(n-2)$
degree in the
Riemann tensor $\hat{R}_{kl}^{ij}$ and the antisymmetrized Kronecker delta $%
\delta _{\lbrack kl]}^{[ij]}$%

\begin{equation}
\mathcal{P}_{j_{5}...j_{2n}}^{i_{5}...i_{2n}}=\sum_{p=0}^{n-2}\frac{D_{p}}{%
\ell ^{2p}}\hat{R}_{j_{5}j_{6}}^{i_{5}i_{6}}...\hat{R}%
_{j_{2(n-p)-1}j_{2(n-p)}}^{i_{2(n-p)-1}i_{2(n-p)}}\delta _{\lbrack
j_{2(n-p)+1}j_{2(n-p+1)}]}^{[i_{2(n-p)+1}i_{2(n-p+1)}]}...\delta
_{\lbrack j_{2n-1}j_{2n}]}^{[i_{2n-1}i_{2n}]},
\end{equation}%
with the coefficients of the expansion given by%

\begin{equation}
D_{p}=\sum_{q=0}^{p}\frac{(-1)^{p-q}}{2q+1}\binom{n-1}{q}.
\end{equation}%

Logically, using the identities for the antisymmetrized Kronecker
deltas, one could express $\mathcal{P}$ in terms of the Riemann
tensor only, but prefer the above form to make easier the
connection with the explicit cases
developed so far. As an example, in nine dimensions, the charge $q(\xi )$ is%

\begin{equation}
q(\xi )=c_{8}\int_{\partial \Sigma }\sqrt{-h}\epsilon
_{i_{1}...i_{8}}(\xi
^{k}K_{k}^{i_{1}})\delta _{j_{2}}^{i_{2}}\Big(\hat{R}%
_{j_{3}j_{4}}^{i_{3}i_{4}}+\frac{1}{\ell ^{2}}\delta _{\lbrack
j_{3}j_{4}]}^{[i_{3}i_{4}]}\Big)\left[ \hat{R}_{j_{5}j_{6}}^{i_{5}i_{6}}\hat{%
R}_{j_{7}j_{8}}^{i_{7}i_{8}}+\frac{3}{5\ell ^{4}}\delta _{\lbrack
j_{5}j_{6}]}^{[i_{5}i_{6}]}\delta _{\lbrack j_{7}j_{8}]}^{[i_{7}i_{8}]}%
\right] dx^{j_{2}}...dx^{j_{8}}
\end{equation}%
whereas in eleven dimensions%

\begin{eqnarray}
q(\xi ) &=&\frac{5c_{10}}{8}\int_{\partial \Sigma
}\sqrt{-h}\epsilon
_{i_{1}...i_{10}}(\xi ^{k}K_{k}^{i_{1}})\delta _{j_{2}}^{i_{2}}\Big(\hat{R}%
_{j_{3}j_{4}}^{i_{3}i_{4}}+\frac{1}{\ell ^{2}}\delta _{\lbrack
j_{3}j_{4}]}^{[i_{3}i_{4}]}\Big)\left[ \hat{R}_{j_{5}j_{6}}^{i_{5}i_{6}}\hat{%
R}_{j_{7}j_{8}}^{i_{7}i_{8}}\hat{R}_{j_{9}j_{10}}^{i_{9}i_{10}}+\right.
\notag \\
&&\!\!\!\!\!\!\!\!\!\!\!\!\!\!\!\!\!\!\!\!\!\!\!\left. +\frac{1}{3\ell ^{2}}\hat{R}_{j_{5}j_{6}}^{i_{5}i_{6}}\hat{R}%
_{j_{7}j_{8}}^{i_{7}i_{8}}\delta _{\lbrack j_{9}j_{10}]}^{[i_{9}i_{10}]}+%
\frac{13}{15\ell ^{4}}\hat{R}_{j_{5}j_{6}}^{i_{5}i_{6}}\delta
_{\lbrack
j_{7}j_{8}]}^{[i_{7}i_{8}]}\delta _{\lbrack j_{9}j_{10}]}^{[i_{9}i_{10}]}-%
\frac{31}{105\ell ^{6}}\delta _{\lbrack
j_{5}j_{6}]}^{[i_{5}i_{6}]}\delta _{\lbrack
j_{7}j_{8}]}^{[i_{7}i_{8}]}\delta _{\lbrack
j_{9}j_{10}]}^{[i_{9}i_{10}]}\right] dx^{j_{2}}...dx^{j_{10}}\!.
\end{eqnarray}%

As a consequence, $q(\xi )$ always vanishes for a spacetime that
is globally AdS. This argument indicates that $q_{0}(\xi )$ in
eq.(\ref{q0tensor2n+1}) for a timelike Killing vector \ is indeed
a covariant formula for the vacuum energy in AAdS spacetimes.

The expression for the vacuum energy has been also recognized as the
action of a Killing vector in the Euclidean continuation of the
boundary term $B_{2n}$ for explicit black hole solutions. It is
expected that a generic thermodynamical relation $S=\beta
(E+E_{0}-\sum_{i}\Omega _{i}J_{i})-I_{2n+1}^{E}=Area/4G$ holds for
any AAdS spacetime that accepts a timelike and a set of rotational
Killing vectors.  Carrying out a similar procedure as Wald's
formalism \cite{wald}, expressions for $E$ and $E_{0}$ should be
mapped exactly to contributions from the bulk and the boundary after
acting with a global isometry
$\xi=\partial_{t}+\Omega^{\infty}_{i}\partial_{\phi_{i}}$ on them,
where the angular velocities at infinity are given by
$\Omega^{\infty}_{i}=a_{i}^{2}/\ell^{2}$.

In odd-dimensional spacetimes, the regularized action is not
invariant under the full AdS group. Radial bulk diffeomorphisms,
which generate a Weyl transformation on the boundary, are
generically broken by the conformal anomaly. In the standard
counterterms approach, this is reflected in a nonvanishing trace
of the regularized stress tensor \cite{Sk-He,Sk,N-O}.

In the present framework, the surface term from an arbitrary
variation of the action contains also variations of the extrinsic
curvature (usually canceled by the Gibbons-Hawking term), such that
a boundary stress tensor definition is not straightforward. The
answer to this point should come from direct comparison of the
Kounterterms series with Dirichlet counterterms, e.g., by expansion
of the tensorial quantities in FG form. For instance, in this way,
it can be proved that in three-dimensional AdS gravity, the
Kounterterms prescription reduces to the Dirichlet regularization up
to a topological invariant on the boundary \cite{Mis-Ole}. The
matching of the results in this paper with the ones obtained by
standard holographic renormalization indicates that both procedures
could also be equivalent in higher dimensions.

If a relation of the Noether charges to a regularized stress tensor
$\tau_{ij}$ is possible, we might expect that a similar splitting as
(\ref{Q2n+1}) would also appear on it. At a more speculative level,
such identification could reveal a connection between the part of
$\tau_{ij}$ that generates the vacuum energy and the one that
produces the Weyl anomaly, what has not yet been understood in the
standard holographic renormalization.

Could the term $B_{d}$ represent the full counterterms series? We
do not know yet the answer to this question. One could only argue
that is very unlikely that other boundary terms can be added on
top of the Kounterterms series, that still preserve the AAdS
boundary conditions extensively used here. This argument might
free this procedure from the ambiguities of the Dirichlet
regularization \cite{NOpred}.

In any case, what is particularly appealing in this formulation is
the relation of Kounterterms to topological invariants in $D=2n$
\cite{OleaJHEP} and to Chern-Simons-like forms (transgression forms) in $D=2n+1$ \cite%
{TFMOTZ}, which might provide some further insight on the problem
of regularization of Einstein-Hilbert-AdS gravity, but also in
Einstein-Gauss-Bonnet \cite{Kof-Ole} and other theories with
higher curvature terms (see, e.g., \cite{MOTZCS,BOTCS}).



\section*{Acknowledgments}

I wish to thank M. Ba\~{n}ados, G. Kofinas, O. Mi\v{s}kovi\'{c},
P. Mora and S. Theisen for helpful discussions. This work was
funded by Funda\c{c}\~{a}o para a Ci\^{e}ncia e Tecnologia (FCT)
of the Ministry of Science, Portugal, through project POCTI/
FNU/44648/2002.



\appendix

\section{Invariant polynomials, Chern-Simons and transgression forms \label{KTTF}}

In this Appendix we review the notion of transgression form as the
natural extension of a Chern-Simons density that restores gauge
invariance through the introduction of an additional gauge
connection \cite{stora,zumino,manes,alvarez}.

Let us consider in $2n+2$ dimensions an invariant polynomial
$~P(F)$~ of the form
\begin{equation}
P(F)=<F^{n+1}> \label{PF}%
\end{equation}
where $F=\frac{1}{2}F_{\mu\nu}^{I}T_{I}\,dx^{\mu}dx^{\nu}=dA+A\wedge
A$ is the
curvature two-form associated to the gauge potential $A=A_{\mu}^{I}%
T_{I}\,dx^{\mu}$ of the Lie group $G$, with a generators set
$\{T_{I}\}$. The symbol $<\dots>$ stands for a totally symmetric
invariant trace of the generators in the adjoint representation of
$G$
\begin{equation}
<T_{I_{1}}\dots T_{I_{n+1}}>=g_{I_{1}\cdots I_{n+1}}.
\end{equation}

The invariant polynomial (\ref{PF}) is a closed form
\begin{equation}
dP(F)=0
\end{equation}
and therefore, by virtue of the Poincar\'{e} lemma, locally exact
\begin{equation}
P(F)=d\mathcal{C}_{2n+1}(A,F)
\end{equation}
what provides the definition of a Chern-Simons density as the
integration over a continuous parameter $u$
\begin{equation}
\mathcal{C}_{2n+1}(A,F)\equiv(n+1)\int_{0}^{1}du~\,<AF_{u}^{n}>
\end{equation}
with $A_{u}=uA$ and $F_{u}=dA_{u}+A_{u}^{2}$.

A similar relation defines a transgression form
$\mathcal{T}_{2n+1}(A,\bar {A})$, that involves two gauge potentials
$A$ and $\bar{A}$ in the same homotopy class, with curvatures $F$
and $\bar{F}$, respectively
\begin{equation}
<F^{n+1}>-<\bar{F}^{n+1}>=d\mathcal{T}_{2n+1}(A,\bar{A}). \label{defTF}%
\end{equation}
The explicit formula for the transgression form is also given by a
parametric integration
\begin{equation}
\mathcal{T}_{2n+1}(A,\bar{A})\equiv(n+1)\int_{0}^{1}dt~\,<(A-\bar{A})F_{t}%
^{n}>, \label{TFAbarA}%
\end{equation}
where $F_{t}=dA_{t}+A_{t}^{2}$ is the curvature associated to the
interpolating gauge connection $A_{t}=tA+(1-t)\bar{A}$. On the
contrary to Chern-Simons densities, transgression forms are truly
invariant under finite gauge transformations in the group $G$.

The explicit formula for (\ref{TFAbarA}) is a consequence of the use
of the \textit{Cartan homotopy operator} $k_{01}$, which acts on
generic polynomials $P(F_{t},A_{t})$ and is defined as
\begin{equation}
k_{01}P(F_{t},A_{t})=\int_{0}^{1}dt~l_{t}P(F_{t},A_{t}),
\end{equation} where the action of the operator $l_{t}$ on arbitrary
polynomials of $A_{t}$ and $F_{t}$ can be worked out from the
relations
\begin{equation}
l_{t}A_{t}=0~~,~~~~l_{t}F_{t}=A-\bar{A}.
\end{equation}
The operator $l_{t}$ acts as an antiderivative
$l_{t}(\Lambda_{p}\Sigma
_{q})=(l_{t}\Lambda_{p})\Sigma_{q}+(-1)^{p}\Lambda_{p}(l_{t}\Sigma_{q})$,
where $\Lambda_{p}$ and $\Sigma_{q}$ are $p$ and $q$-forms,
respectively.

It is particularly useful to express the curvature $F_{t}$ as%

\begin{equation}
F_{t}=\bar{F}+t\bar{D}(A-\bar{A})+t^{2}(A-\bar{A})^{2},
\end{equation}
with the curvature $\bar{F}$ associated to the connection $\bar{A}$
($\bar {F}=d\bar{A}+\bar{A}^{2}$) and the covariant derivative in
$\bar{A}$ given by
$\bar{D}(A-\bar{A})=d(A-\bar{A})+\bar{A}(A-\bar{A})+(A-\bar{A})\bar{A}$.

In four dimensions, one can re-obtain the formula for the second
Chern form (\ref{B3CF}) from the generic transgression formula
taking two gauge connections for the Lorentz group $SO(3,1)$, that
is, $A=\frac{1}{2}\omega^{AB}J_{AB}$ and
$\bar{A}=\frac{1}{2}\bar{\omega}^{AB}J_{AB}$ and the invariant
tensor for the Lorentz generators
$\{J_{AB}J_{CD}\}=\varepsilon_{ABCD}$. In this way, the
interpolating connection in terms of the Second Fundamental Form
(\ref{SFFdef}) is
\begin{equation}
A_{t}=\frac{1}{2}\omega_{t}^{AB}J_{AB}=\frac{1}{2}(\bar{\omega}^{AB}%
+t\,\theta^{AB})J_{AB},
\end{equation}
and its corresponding curvature%

\begin{equation}
F_{t}=\frac{1}{2}\hat{R}_{t}^{AB}J_{AB}=\frac{1}{2}\left[  \hat{\bar{R}%
}^{AB}+t\,\bar{D}\,\theta^{AB}+t^{2}\theta_{\,C}^{A}\theta^{CB}\right]
J_{AB},
\end{equation}
where $\hat{\bar{R}}^{AB}$ and $\bar{D}$ are the curvature and the
covariant derivative in the spin connection $\bar{\omega}^{AB}$.
When plugged in Eqs.(\ref{defTF},\ref{TFAbarA}), the transgression
form for the Lorentz group satisfies the local
relation%
\begin{equation}
\mathcal{E}_{4}(\hat{R})-\mathcal{E}_{4}(\hat{\bar{R}})=2d\left(
\int\limits_{0}^{1}dt\,\varepsilon_{ABCD}\,\theta^{AB}\left( \hat
{\bar{R}}^{CD}+t\,\bar{D}\,\theta^{CD}+t^{2}\theta_{\,F}^{C}\theta
^{FD}\right)  \right)  , \label{Eu-Eu}%
\end{equation}
where $\mathcal{E}_{4}=\varepsilon_{ABCD}\hat{R}^{AB}\hat{R}^{CD}$
is the Euler-Gauss-Bonnet topological invariant. For a radial
foliation of the spacetime (\ref{Ncoord}), an adequate choice of the
reference spin connection corresponds to the matching conditions
(\ref{SFFnortan}) for the Second Fundamental Form and relates the
components $\hat{\bar{R}}^{ab}$ to the intrinsic curvature at the
boundary, i.e., $\hat{\bar{R}}^{ab}=R^{ab}(h)$. In doing so, the
second Euler term in the l.h.s. of Eq.(\ref{Eu-Eu}) vanishes
identically, so does the second term in the r.h.s.

Global considerations show that the Euler term and the second Chern
form $B_{3}$ (\ref{B3CF}) in four dimensions are equivalent up a
topological number (Euler
characteristic $\chi(M_{4})$)%
\begin{equation}
\int\limits_{M_{4}}\mathcal{E}_{4}(\hat{R})=32\pi^{2}\chi(M_{4})+\int
\limits_{\partial M_{4}}B_{3},
\end{equation}
revealing the profound connection of the Kounterterms method with
topological invariants.

In a similar fashion, the higher even-dimensional Kounterterms
$B_{2n-1}$ (\ref{B2ntheta}) are related to the corresponding Euler
term in $D=2n$ dimensions by virtue of
the Euler theorem%

\begin{equation}
\int\limits_{M_{2n}}\mathcal{E}_{2n}(\hat{R})=\left(  -4\pi\right)
^{n}\,n!\,\chi(M_{2n})+\int\limits_{\partial M_{2n}}B_{2n-1}.
\end{equation}

For any invariant polynomial $P(F_{t},A_{t})$, it can be verified
that
\begin{equation}
\big(l_{t}d+dl_{t}\big)P(F_{t},A_{t})=\frac{\partial}{\partial t}P(F_{t}%
,A_{t})
\end{equation} which, when integrated between 0 and 1,
recovers the Cartan homotopy formula
\begin{equation}
\big(k_{01}d+dk_{01}\big)P(F_{t},A_{t})=P(F,A)-P(\bar{F},\bar{A})~.
\end{equation}
In particular, for $P=\mathcal{C}_{2n+1}$, the above relation allows
us to express a transgression form as the difference of two
Chern-Simons densities plus a boundary term
\begin{equation}
\mathcal{T}_{2n+1}=\mathcal{C}_{2n+1}(A,F)-\mathcal{C}_{2n+1}(\bar{A},\bar
{F})+d\Xi_{2n}(A,F;\bar{A},\bar{F}). \label{TF-CS}%
\end{equation}
The $2n-$form $\Xi_{2n}$ is defined by the action of the Cartan
homotopy operator on a Chern-Simons term
\begin{equation}
\Xi_{2n}(A,F;\bar{A},\bar{F})\equiv k_{01}\mathcal{C}_{2n+1}%
\end{equation} whose explicit form is given by
\begin{equation}
\Xi_{2n}=n(n+1)\int_{0}^{1}ds\int_{0}^{1}dt~s~<A_{t}(A-\bar{A})~F_{st}^{n-1}>
\label{Xi2n}%
\end{equation}
where $F_{st}=sF_{t}+s(s-1)A_{t}^{2}$.

When the gauge group is AdS, the connection field is written as%

\begin{equation}
A=\frac{1}{2}\omega^{AB}J_{AB}+e^{A}P_{A},
\end{equation} where
$J_{AB}$ and $P_{A}$ are the generators of rotations and AdS
translations, respectively. For the trace of the generators of
$SO(2n,2)$, we
take $<J_{A_{1}A_{2}},...,J_{A_{2n-1}A_{2n}},P_{A_{2n+1}}>=\frac{2^{n}}%
{(n+1)}\varepsilon_{A_{1}\cdots A_{2n+1}}$, such that each
Chern-Simons form in (\ref{TF-CS}) can be written as a bulk Lovelock
Lagrangian (a polynomial in
the curvature and the vielbein)%
\begin{eqnarray}
\mathcal{C}_{2n+1}(A,F) &=& \mathcal{L}_{CS-AdS}(\hat{R},e) \nonumber \\
&=& \!\!\!\!\int_{0}%
^{1}\!\!dt\,\varepsilon_{A_{1}\cdots A_{2n+1}}\!(\hat{R}^{A_{1}A_{2}}\!+\!\frac{t^{2}%
}{\ell^{2}}e^{A_{1}}e^{A_{2}})\!\!\cdots \!\!(\hat{R}^{A_{2n-1}A_{2n}}\!+\!\frac{t^{2}%
}{\ell^{2}}e^{A_{2n-1}}e^{A_{2n}})e^{A_{2n+1}}
\end{eqnarray}
plus a surface term whose explicit form is involved and has to be
worked out case by case. However, the matching conditions for the
Second Fundamental Form (\ref{SFFnortan}) --that single out the
boundary correction in the Euler theorem-- and the condition
$\bar{e}^{A}=0$ (in order to avoid a background-dependent bimetric
formulation) replaced in the transgression form (\ref{TF-CS}) result
into a gravity action
\begin{equation}
I_{CS-AdS} = \int\limits_{M_{2n+1}}\mathcal{T}_{2n+1}\\
= \!\!\int\limits_{M_{2n+1}%
}\mathcal{L}_{CS-AdS}(\hat{R},e)+\int\limits_{\partial M_{2n+1}}B_{2n}%
(\theta,e)
\end{equation}
with a boundary term $B_{2n}(\theta,e)$ given by the formula
(\ref{B2ntheta}). The above total action  is regularized both in its
Euclidean continuation and in the Noether charges. It is quite
remarkable that the contributions from the bulk terms in
(\ref{TF-CS}) combine to the surface term (\ref{Xi2n}) to produce a
compact expression for the boundary terms in this particular
Lovelock theory as the double integral in Eq.(\ref{B2ntheta}). But
it is even more surprising that the same $2n-$form is useful to
regularize other gravity theories, including Einstein-Hilbert-AdS,
what is shown explicitly in this paper.

\section{Useful Identities \label{UI}}

Let us consider the five-dimensional Kounterterms as an example of
the equivalence between differential forms and tensorial notation

\begin{eqnarray}
B_{4} &=&\epsilon _{A_{1}...A_{5}}\theta ^{A_{1}A_{2}}e^{A_{3}}\Big(%
R^{A_{4}A_{5}}+\frac{1}{2}\theta _{\,\,\,\,C}^{A_{4}}\theta ^{CA_{5}}+\frac{1%
}{6\ell ^{2}}e^{A_{4}}e^{A_{5}}\Big), \notag \\
&=&2\varepsilon _{1a_{1}a_{2}a_{3}a_{4}}\theta ^{1a_{1}}e^{a_{2}}\Big(%
R^{a_{3}a_{4}}+\frac{1}{2}\theta _{\,\,\,\,1}^{a_{3}}\theta ^{1a_{4}}+\frac{1%
}{6\ell ^{2}}e^{a_{3}}e^{a_{4}}\Big), \notag \\
&=&-2\varepsilon _{a_{1}a_{2}a_{3}a_{4}}K^{b}e^{c}\Big(R^{a_{3}a_{4}}-\frac{1%
}{2}K^{a_{3}}K^{a_{4}}+\frac{1}{6\ell ^{2}}e^{a_{3}}e^{a_{4}}\Big), \notag \\
&=&-\varepsilon
_{a_{1}a_{2}a_{3}a_{4}}e_{j_{1}}^{a_{1}}e_{j_{2}}^{a_{2}}e_{j_{3}}^{a_{3}}e_{j_{4}}^{a_{4}}K_{i_{1}}^{j_{1}}\delta _{i_{2}}^{j_{2}}%
\Big(R_{i_{3}i_{4}}^{j_{3}j_{4}}-\frac{1}{2}K_{[i_{3}i_{4}]}^{[j_{3}j_{4}]}+%
\frac{1}{6\ell ^{2}}\delta _{\lbrack i_{3}i_{4}]}^{[j_{3}j_{4}]}\Big)%
dx^{i_{1}}\wedge ...\wedge dx^{i_{4}},\notag \\
&=&\sqrt{-h}\delta _{\lbrack
j_{1}j_{2}j_{3}]}^{[i_{1}i_{2}i_{3}]}K_{i_{1}}^{j_{1}}\Big(%
R_{i_{2}i_{3}}^{j_{2}j_{3}}-K_{i_{2}}^{j_{2}}K_{i_{3}}^{j_{3}}+\frac{1}{%
3\ell ^{2}}\delta
_{i_{2}}^{j_{2}}\delta_{i_{3}}^{j_{3}}\Big)d^{4}x,
\end{eqnarray}%
where we have used the identities%

\begin{equation}
\varepsilon _{a_{1}...a_{2n}}e_{j_{1}}^{a_{1}}...e_{j_{2n}}^{a_{2n}}=-\sqrt{%
-h}\varepsilon _{j_{1}...j_{2n}},
\end{equation}%
with the definition $\sqrt{-h}=\det (e)$, the volume element

\begin{equation}
dx^{i_{1}}\wedge ...\wedge dx^{i_{2n}}=\varepsilon
^{i_{1}...i_{2n}}d^{2n}x,
\end{equation}%
and the general property for antisymmetrized Kronecker deltas%

\begin{equation}
\delta _{\lbrack j_{1}...j_{p}]}^{[i_{1}...i_{p}]}\delta
_{i_{1}}^{j_{1}}\delta _{i_{2}}^{j_{2}}...\delta _{i_{m}}^{j_{m}}=\frac{%
(r-p+m)!}{(r-p)!}\delta _{\lbrack
j_{m+1}...j_{p}]}^{[i_{m+1}...i_{p}]}
\end{equation}%
where $p>m$ and $r$ as the range of the indices.

\section{Noether's Theorem \label{Noether}}

We now recall the standard construction of the conserved
quantities associated to asymptotic symmetries of the action
through the Noether's theorem.

Let us consider an action that is the integral of a $D-$form
Lagrangian density in $D$ dimensions

\begin{equation}
L=\frac{1}{D!}L_{\mu _{1}...\mu _{D}}dx^{\mu _{1}}\wedge ...\wedge
dx^{\mu _{D}}.  \label{Ldform}
\end{equation}

An arbitrary variation $\bar{\delta}$ acting on the fields can be
always decomposed in a functional variation $\delta $ plus the
variation due to an infinitesimal change in the coordinates
$x^{\prime \mu }=x^{\mu }+\eta ^{\mu }$. For a $p$-form field
$\varphi $, the latter variation is given by the
Lie derivative $\mathcal{L}_{\eta }\varphi $ along the vector $\eta ^{\mu }$%
, that can be written as $\mathcal{L}_{\eta }\varphi =(dI_{\eta
}+I_{\eta }d)\varphi $, where $d$ is the exterior derivative and
$I_{\eta }$ is the contraction operator \footnote{%
The action of the contraction operator $I_{\eta }$ over a $p$-form
$\alpha _{p}=\frac{1}{p!}\alpha _{\mu _{1}}\ldots _{\mu
_{p}}dx^{\mu_{1}}\wedge
\ldots \wedge dx^{\mu _{p}}$ is given by $I_{\eta }\alpha _{p}=\frac{1}{%
(p-1)!}\eta ^{\nu }\alpha _{\nu \mu _{1}}\ldots _{\mu
_{p-1}}dx^{\mu
_{1}}\wedge \ldots \wedge dx^{\mu _{p-1}}$}. The functional variation $%
\delta $ acting on $L$ produces the equations of motion plus a surface term $%
\Theta (\varphi ,\delta \varphi )$. The Lie derivative contributes
only with another surface term because $dL=0$ in $D$ dimensions.

The Noether's theorem provides a conserved current associated to
the invariance under diffeomorphisms of the Lagrangian $L$, that
is given by \cite{Choquet-Dewitt,Ramond}

\begin{equation}
\ast J=-\Theta (\varphi ,\delta \varphi )-i_{\eta }L.
\label{Jdiff}
\end{equation}%

In case that the diffeomorphism $\xi $ is a Killing vector, we
have $\delta \varphi =-\mathcal{L}_{\xi }\varphi $, with
$\mathcal{L}_{\xi }$ the Lie derivative along the vector $\xi
^{\mu }$. Because the surface term $\Theta $
is linear in the variations of the fields, the current takes the form%

\begin{equation}
\ast J=\Theta (\varphi ,\mathcal{L}_{\xi }\varphi )-i_{\xi }L.
\label{JKill}
\end{equation}%

(see also \cite{Iyer-Wald} and, for a recent discussion, \cite{H-I-M,HIM2}).

As the current is conserved ($d\ast J=0$), $\ast J$ can always be
written locally as an exterior derivative of a quantity. In general, the boundary $%
\partial M$ consists of two spacelike surfaces (at initial time $\Sigma
_{t_{1}}$ and at final time $\Sigma _{t_{2}}$) and a timelike surface $%
\Sigma _{\infty }$ (at spatial infinity). Only when the current
can be written globally as an exact form $\ast J=dQ(\xi )$, we can
integrate the charge $Q(\xi )$ in a $(D-2)-$dimensional surface
$\partial \Sigma $ (a constant-time slice in $\Sigma _{\infty }$,
as we assume flux conservation through $\Sigma _{t_{1}}$ and
$\Sigma _{t_{2}}$).

Let us consider now a Lagrangian $\mathcal{L}$ that differs from $L$ in a
boundary term $d\beta $%

\begin{equation}
\mathcal{L}=L+dB,
\end{equation}%
so that the conserved current is modified as%

\begin{eqnarray}
\mathcal{\ast J} &=&\Theta (\varphi ,\mathcal{L}_{\xi }\varphi )-i_{\xi }L+%
\frac{\delta B}{\delta \varphi }\mathcal{L}_{\xi }\varphi -i_{\xi
}dB
\label{Jnew} \\
&=&d\left( Q(\xi )+i_{\xi }B\right) .  \label{JQnew}
\end{eqnarray}%

The above formula provides a useful shortcut to find the conserved charges
of a Lagrangian supplemented by a boundary term as%

\begin{equation}
\mathcal{Q}(\xi )=Q(\xi )+i_{\xi }B.  \label{Qnew}
\end{equation}

\section{Variation of $B_{2n}$ \label{varB2n}}

The variation of the Kounterterms series $B_{2n}$ has an expanded
form given by

\begin{eqnarray}
\delta B_{2n} &=&-2n\int\limits_{0}^{1}dt\varepsilon
\sum\limits_{k=0}^{n-1}C_{k}^{n-1}\sum%
\limits_{l=0}^{n-1-k}C_{l}^{n-1-k}R^{n-1-k-l}\int%
\limits_{0}^{t}dst^{2k+1}(-1)^{k}\delta K^{2k+1}s^{2l}e^{2l+1}  \notag \\
\!\!\!\!\!&&-2n\int\limits_{0}^{1}dt\varepsilon
\sum\limits_{k=0}^{n-1}C_{k}^{n-1}\sum%
\limits_{l=0}^{n-1-k}C_{l}^{n-1-k}R^{n-1-k-l}\int%
\limits_{0}^{t}dst^{2l+1}(-1)^{l}K^{2l+1}s^{2k}\delta
e^{2k+1}\!\!,
\end{eqnarray}%
where variations of the intrinsic curvature produce surface terms
that are identically vanishing on the boundary. After some
algebraic manipulations, we have

\begin{eqnarray}
\delta B_{2n} &=&-2n\int\limits_{0}^{1}dt\varepsilon \delta
Ke\sum\limits_{k=0}^{n-1}C_{k}^{n-1}\left( R+t^{2}ee\right)
^{n-1-k}(-KK)^{k}\left( 1-t^{2k+1}\right)  \notag \\
&&-2n\int\limits_{0}^{1}dt\varepsilon K\delta
e\sum\limits_{k=0}^{n-1}C_{k}^{n-1}\left( R+t^{2}e^{2}\right)
^{n-1-k}t^{2k+1}(-KK)^{k},
\end{eqnarray}%
or, in a more convenient form%

\begin{eqnarray}
\delta B_{2n} &=&-2n\int\limits_{0}^{1}dt\varepsilon \delta
Ke\left(
R-KK+t^{2}ee\right) ^{n-1}  \notag \\
&&+2n\int\limits_{0}^{1}dt\varepsilon (\delta Ke-K\delta e)\left(
R-t^{2}KK+t^{2}e^{2}\right) ^{n-1},
\end{eqnarray}%
that is particularly useful to impose the asymptotic conditions
for AAdS spacetimes.



\end{document}